\documentclass[pre,twocolumn,superscriptaddress]{revtex4}

\usepackage{graphicx,xcolor}
\usepackage{amsmath,amsfonts,amssymb,bm}
\usepackage{listings}
\usepackage{soul}

\usepackage[
unicode=true,
bookmarks=true,%
bookmarksnumbered=true,%
hyperfootnotes=false,
colorlinks=true,%
linkcolor=blue,%
citecolor=blue,%
urlcolor=blue
]{hyperref}  

\DeclareMathOperator{\sech}{sech}

\newenvironment{mat}{\left[\begin{array}{ccccccccccccccc}}{\end{array}\right]}
\newcommand\bcm{\begin{mat}}
\newcommand\ecm{\end{mat}}
\newcommand\pFq[2]{{}_{#1}F_{#2}}

\newenvironment{cmat}{\left(\begin{array}{ccccccccccccccc}}{\end{array}\right)}
\newcommand\bcrm{\begin{cmat}}
\newcommand\ecrm{\end{cmat}}

\newenvironment{rmat}{\left[\begin{array}{rrrrrrrrrrrrr}}{\end{array}\right]}
\newcommand\brm{\begin{rmat}}
\newcommand\erm{\end{rmat}}

\definecolor{blue1}{RGB}{90, 200, 255}
\definecolor{pink1}{RGB}{255, 153, 153}
\definecolor{green1}{RGB}{56, 210, 92}

\begin{document}

\title{Rogue and solitary waves in coupled phononic crystals}

\author{Y. Miyazawa}
\affiliation{Department of Aeronautics and Astronautics, University of Washington, Seattle, Washington 98195}
\author{C. Chong}
\affiliation{Department of Mathematics, Bowdoin College, Brunswick, Maine 04011}
\author{P. G. Kevrekidis}
\affiliation{Department of Mathematics and Statistics, University of Massachusetts Amherst, Amherst, Massachusetts 01003-4515}
\author{J. Yang}
\affiliation{Department of Aeronautics and Astronautics, University of Washington, Seattle, Washington 98195}
\date{\today}

\begin{abstract}
In this work we present an analytical and numerical study of rogue and solitary waves in a coupled one-dimensional nonlinear lattice that involves both axial and rotational degrees of freedom.
Using a multiple-scale analysis we derive a system of coupled
nonlinear Schr\"odinger type equations in order to approximate
solitary waves and rogue waves of the coupled lattice model.
Numerical simulations are found to agree with the analytical approximations.
We also consider generic initialization data in the form of a Gaussian profile and observe that they can result in the spontaneous formation of rogue wave-like patterns in the lattice.
The solitary and rogue waves in the lattice demonstrate both energy isolation and exchange between the axial and rotational degrees of freedom of the system.
This suggests that the studied coupled lattice has the potential to be an efficient energy isolation, transfer, and focusing medium.
\end{abstract}

\maketitle

\section{Introduction}
Rogue waves are large waves which appear suddenly and disappear without a trace~\cite{Kharif2003}.
Ocean rogue waves were first measured in the North Sea several decades ago~\cite{Haver2004,Walker2004,Adcock2011,Mori2002} sparking interest in their scientific study. Since then, multiple measurements were conducted elsewhere across the globe~\cite{Mori2002,Baschek2011,Pelinovsky2016}, providing evidence that ocean rogue waves are an important feature worthy of further
exploration.
According to the statistical maritime definition, rogue waves are localized both in space and time with an amplitude two times larger than the significant wave height~\mbox{\cite{Kharif2003}}.
The study of the rogue wave has gone well beyond oceanographic settings, and includes other spatially continuous systems, such as water 
tanks~\cite{Chabchoub2011,Chabchoub2012,Adcock2018,Chabchoub2020}, ultra-cold bosonic gases~\cite{Charalampidis2018a}, nonlinear optics~\cite{Solli2007,dudley1,Kibler2016,Tikan}, microwave transport~\cite{Hohmann2010}, and space plasma~\cite{Ruderman2010,Sabry2012,Bains2014,Tolba2015}.  
Indeed, at this point, numerous reviews~\cite{Onorato2013} and books~\cite{R.Osborne2010,Pelinovsky2016} have summarized the 
rapidly expanding state-of-the-art on the subject.

A central possibility towards the existence of rogue waves, including many of the above themes, involves the nonlinear affects of the underlying system.
In particular, in the previously mentioned physical settings the focusing nonlinear Schr\"odinger equation~(NLSE)~\cite{Sulem1999,Ablowitz2003} can be derived as an approximate
model under a suitable set of assumptions/approximations. Among the exact solutions of NLSE, the Peregrine soliton solution~\cite{Peregrine1983} is considered a prototypical example of a rogue wave, given that it has only one localized peak in the spatio-temporal domain. The peak amplitude is three times larger than the background plane wave amplitude and satisfies the classical maritime definition of the rogue wave.
Indeed, not only the Peregrine soliton, but even the corresponding
higher order (breather) generalizations thereof have been observed
in recent experiments~\cite{Chabchoub2012}.

Despite the vast amount of recent activity on the study of rogue waves, there have been relatively few reports on their study in solids or structures
and more concretely in associated spatially discrete models.
Only recently, rogue waves in chains of interacting particles (so-called granular crystals) have been numerically and analytically explored~\cite{Charalampidis2018}.
Another example of a discrete setting where rogue waves have been studied is the integrable Ablowitz-Ladik lattice \cite{Akhmediev2011}, which is known to have an exact solution that has similar
properties as the NLSE Peregrine soliton.
Rogue waves have also been studied in the discrete Hirota lattice~\mbox{\cite{Ankiewicz2010,Wen2018}} and Salerno lattice~\mbox{\cite{Yan2012,Maluckov2013}}.
It is interesting to note that in a number of relevant NLSE lattice models, it was recognized that rogue waves are more likely to arise at or near the integrable limit (such as the Ablowitz-Ladik lattice), rather than its non-integrable analogue, e.g., the standard discrete NLSE case~\cite{Maluckov2013,HOFFMANN20183064,sullivan}.

At the level of granular systems, the pioneering work of~\cite{Han2014} was the first one, to our knowledge, to recognize the potential of such systems for unusually large (rogue) fluctuations in late time dynamics, in the absence of dissipation. Recent work in this direction has, in fact, posited that in Fermi-Pasta-Ulam-Tsingou (FPUT) non-integrable lattices, rogue fluctuations may be generic for sufficiently long times~\cite{sen2}.

Models of one-dimensional (1D) lattices that include additional degrees of freedom have also gained significant recent attention~\cite{Merkel,Pichard2014,Kopfler2019,Ngapasare2020,Allein2017,Dubus2016,Deng2018,Zhang2019}.
For example, the standard model of the granular crystal accounts for axial (translational) motion of the particles but ignores any rotation.
Models that account for the additional degree of freedom in the form of rotation have the obvious benefit of being more realistic representations of the physical system, but such models can also lead to other novel dynamics such as rotational-translational modes \cite{Merkel}.
Further studies have demonstrated the localized translational-rotational modes in the coupled linear systems~\mbox{\cite{Pichard2014}}, which can offer mechanisms for energy transfer from one degree-of-freedom to another by utilizing topologically protected modes~\mbox{\cite{Kopfler2019}}.
The wave propagation in the linear, multi degree-of-freedom 1D lattice has also been shown to facilitate the energy spreading~\mbox{\cite{Ngapasare2020}}, or be easily manipulated by tuning the lattice configuration by using one of the degrees of freedom as a control knob in the magneto-granular crystal~\mbox{\cite{Allein2017}}.
By introducing nonlinearity, the linear dispersion relationship can be corrected with nonlinear terms resulting in nonlinear resonances that can significantly enhance the energy harvesting capability of the lattice~\cite{Dubus2016,Zhang2019}. The action of nonlinearity may also
have significant further implications, such as the existence of amplitude gaps for the existence of traveling (nonlinear) waves~\cite{Deng2018} in a metamaterial lattice constructed out of LEGO bricks. 
Another example of a coupled system (which incorporates axial and rotational degrees of freedom) is the origami-inspired mechanical lattice.
Recently, it has been shown that rarefaction solitary waves exist in this lattice~\cite{Yasuda2019}.
Elastic vector solitons with more than two components (e.g., two translational and one rotational) have also been studied recently via combinations of analytical and numerical tools and present a rich phenomenology in their own right, including the potential emergence of focusing, sound-bullet forming events~\cite{Deng2019}.

In the present study, we consider a lattice with two coupled channels (i.e., one that accounts for two degrees of freedom) with a polynomial nonlinearity to explore wave focusing  events, leading to the potential formation of solitary
or of rogue waves.
The coupling mechanism investigated in this study can either facilitate or prevent the transfer of energy between two modes.
For example, we can manage mechanical energy (e.g., energy harvesting, vibration filtering, and impact mitigation) in one mode by imposing a specific initial condition in the other mode.
This control mechanism can be potentially useful for multiple
degree-of-freedom mechanical setups, which are ubiquitous in engineering systems, such as beams~\mbox{\cite{Sugino2017,Beli2018,Karttunen2020}}, plates~\mbox{\cite{Sugino2017}}, tensegrity~\mbox{\cite{Wang2018a,Yin2020,Zhang2021}}, and origami~\mbox{\cite{Yasuda2019,Fang2020,Pratapa2018}}.
The study of such effects on the general coupled nonlinear lattice may, in fact, be 
of broader interest to applications not only in engineering fields such as efficient energy transfer and harvesting, but also in other discrete physics platforms, such as granular crystals in substrates~\mbox{\cite{Zhang2019}} and nonlinear DNA dynamics~\mbox{\cite{Peyrard1989,Dauxois1993,Chevizovich2020}}.

The paper is structured as follows: In Sec.~\ref{model} we introduce the physical set-up and corresponding model equations. An analytical approximation
is derived in Sec.~\ref{ms} by performing a multiple-scale expansion to obtain an NLSE-like system.
 Section~\ref{soliton} summarizes the exact and approximate 
solitary waves of the derived NLSE, which are used to initialize the simulations of the full lattice model,
yielding good agreement between the NLSE-based approximation and the
full direct numerical simulation of the original nonlinear lattice system.
Sec.~\ref{peregrine} considers simulations with initial data given by the Peregrine solution
of the derived NLSE-like system. More general conditions leading to the formation of rogue-wave like structures are considered in
Sec.~\ref{gauss} where simulations starting from (more ``generic'') 
Gaussian initial data are used. The energy exchanged between the
two channels (i.e., the different degrees of freedom) is quantified in Sec.~\ref{exchange}. Section~\ref{theend} concludes
the paper and presents some possible directions for further study.

\section{Physical set-up and mathematical model} \label{model}
\begin{figure}[bp]
    \centering
    \includegraphics[width=\linewidth]{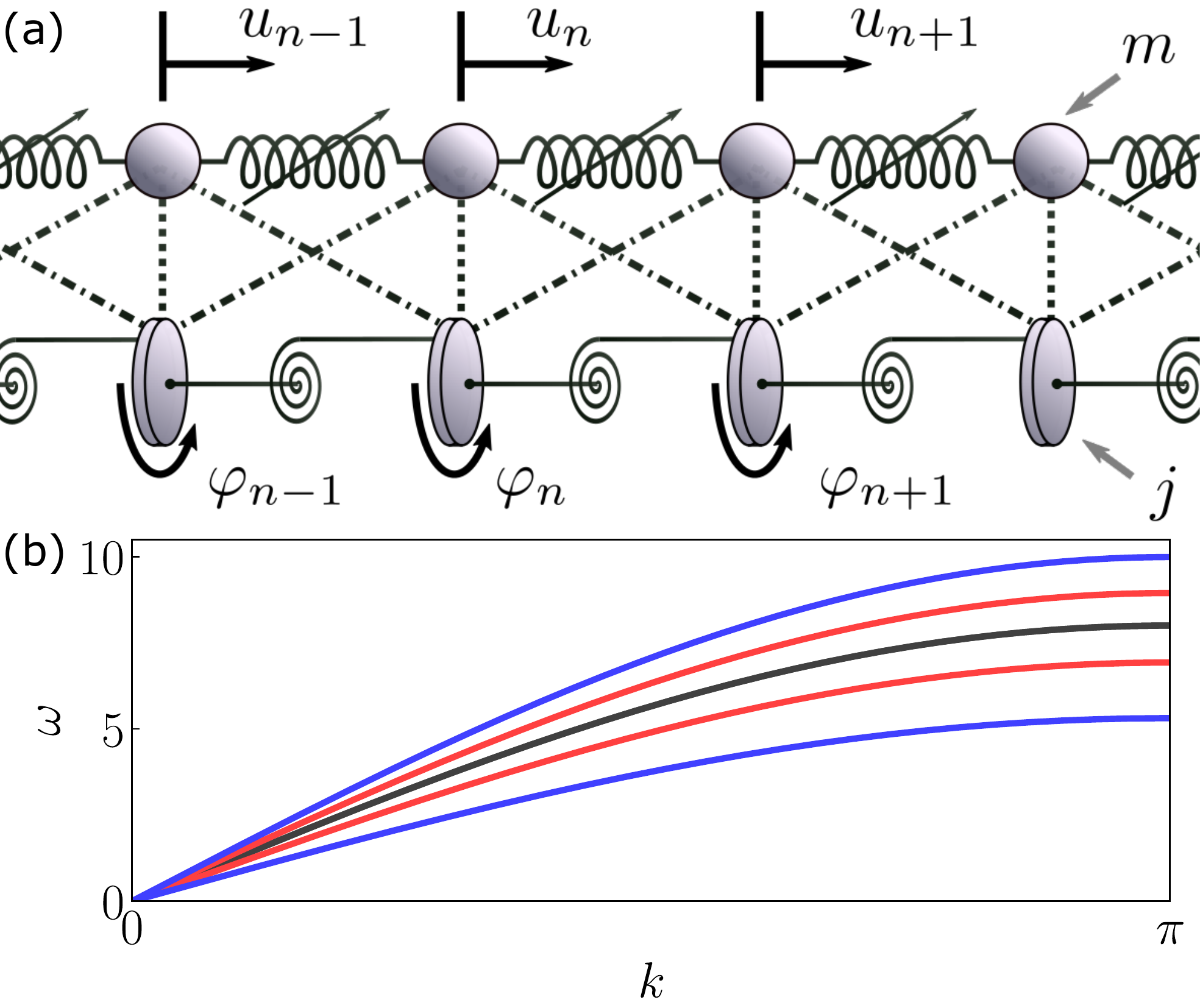}
    \caption{(a) 
    Schematic of 1D lattice with two degrees of freedom modeled as a coupled lattice system with lumped mass $m$ and disc with rotational inertia $j$
    The lumped masses are connected with nonlinear springs, and so are the discs. Adjacent masses and discs are connected with nonlinear springs as well, denoted as dashed and dash-dotted lines.
    (b) The dispersion relationship both with and without the coupling term $\alpha_{12}$. Black dashed line: $\alpha_{11}=\alpha_{22}=16$ and $\alpha_{12}=0$; red solid lines: $\alpha_{11}=20$, $\alpha_{22}=12$, and $\alpha_{12}=0$; blue solid lines: $\alpha_{11}=20$, $\alpha_{22}=12$, and $\alpha_{12}=8$.
    }
    \label{fig:schematic}
\end{figure}

In this study, we consider a lattice consisting of particles with two degrees of freedom: an axial degree of freedom $u$ and a rotational degree of freedom $\varphi$.
The particles have mass $m$ and rotational inertia $j$, and are connected to each other via nonlinear springs.
We model this lattice as the coupled system illustrated in Fig.~\ref{fig:schematic}(a), where axial and rotational degrees of freedom are considered separately.
In this visualization, there are two one-dimensional lattices composed of lumped masses and inertial discs that are connected to each other via nonlinear springs.
The model is mathematically equivalent to a 1D lattice of unit cells where each unit exhibits axial and rotational motion.
Examples where our model would be relevant include the aforementioned granular crystal~\mbox{\cite{Merkel}}, Kresling origami~\mbox{\cite{Kresling2012}}, and a compliant mechanism~\mbox{\cite{Howell2013}}.
Note, if the coupled nonlinear springs are significantly stiff, the lattice can be considered as a quasi-one degree of freedom system \cite{Yasuda2019}.

The coupled lattice is governed by the following equations of motion,
\begin{subequations}\label{eq:eom}
\begin{align}
    m_n\Ddot{u}_n=V_1^{\prime}(\Delta u_{n-1},\Delta\varphi_{n-1})-V_1^{\prime}(\Delta u_n,\Delta\varphi_n),\\
    j_n\Ddot{\varphi}_n=V_2^{\prime}(\Delta u_{n-1},\Delta\varphi_{n-1})-V_2^{\prime}(\Delta u_n,\Delta\varphi_n),
\end{align}
\end{subequations}
where $m_n$ and $j_n$ are the mass and rotational inertia of the axial and rotational component, respectively, $\Delta u_n=u_n-u_{n+1}$ and $\Delta\varphi_n=\varphi_n-\varphi_{n+1}$ are the axial and rotational strains, with $u_n$ and $\varphi_n$ being the axial displacement and angle of rotation of the $n$-th particle respectively.
$V_1^{\prime}$ and $V_2^{\prime}$ are the general nonlinear force and torque terms determined by differentiating the total potential energy $V=V(\Delta u,\Delta\varphi)$ of the unit cell as follows,
\begin{align}
    V^\prime_1=\frac{\partial V}{\partial (\Delta u)},\quad V^\prime_2=\frac{\partial V}{\partial (\Delta\varphi)}.
\end{align}
Here, the total potential energy $V$ is a function of $\Delta u$ and $\Delta\varphi$, and therefore the Hamiltonian of this system is:
\begin{align}
    H=\sum_{n\in\mathbb{Z}}\left[\frac{1}{2}\left(m_n\Dot{u}_n^2+j_n\Dot{\varphi}_n^2\right)+V(\Delta u_n,\Delta\varphi_n)\right].
\end{align}

In the present work, we assume that the masses are identical ($m_n = m$ and $j_n = j$), and that the potential $V$ is a fourth order polynomial, which can be thought of as a Taylor expansion of an application specific potential
(e.g., $V$ has the form of a power-law in the case of the precompressed granular crystal lattice \cite{Nesterenko2001}). In particular, the total  potential energy function $V$ considered here is:
\begin{align}\label{eq:pe_nondim}
    V(x,y)&=\frac{1}{2}\alpha_{11}x^2+\alpha_{12}xy+\frac{1}{2}\alpha_{22}y^2\nonumber\\
                &~~~+\frac{1}{6}\alpha_{111}x^3+\frac{1}{2}\alpha_{112}x^2y
                +\frac{1}{2}\alpha_{122}xy^2+\frac{1}{6}\alpha_{222}y^3\nonumber\\
                &~~~+\frac{1}{24}\alpha_{1111}x^4+\frac{1}{6}\alpha_{1112}x^3y+\frac{1}{4}\alpha_{1122}x^2y^2\nonumber\\
                &~~~+\frac{1}{6}\alpha_{1222}xy^3+\frac{1}{24}\alpha_{2222}y^4.
\end{align}
In the above definition, we assumed non-dimensional parameters (note that we retained the same symbols for $u,~\varphi,~t$)
\begin{align}\label{eq:nondim_param}
    u_n\rightarrow\frac{u_n}{D_0},\quad \varphi_n\rightarrow\frac{R_0\varphi_n}{D_0},\quad t\rightarrow\omega_0t
\end{align}
where $D_0$ is the lattice constant, $R_0$ is the radius of the particle (i.e., of the disc in Fig.~{\ref{fig:schematic}(a)}), and $T_0=1/\omega_0=c\sqrt{m/a_{11}}$ is the characteristic time scale.
The parameter $c$ is an arbitrary real constant such that $\alpha_{11}=c^2$ and can be set to any positive real values including $\alpha_{11}=4$, which we use in the following sections.
The $a_{11}$ is the dimensional linear stiffness coefficient of the axial channel~(see Supplementary Note~1
for how the non-dimensional coefficients $\alpha$ are related to the dimensional coefficients $a$).
With this rescaling, the coupled equations of motion become
\begin{subequations}\label{eq:eom_nondim}
\begin{align}
    \Ddot{u}_n&=V_1^{\prime}(\Delta u_{n-1},\Delta\varphi_{n-1})-V_1^{\prime}(\Delta u_n,\Delta\varphi_n),\\
    \Ddot{\varphi}_n&=V_2^{\prime}(\Delta u_{n-1},\Delta\varphi_{n-1})-V_2^{\prime}(\Delta u_n,\Delta\varphi_n).
\end{align}
\end{subequations}

\section{Multiple-scale expansion} \label{ms}
To analytically explore the behavior of our coupled lattice, we employ asymptotic expansions accompanied with multiple-scale variables~\cite{Huang1993,Charalampidis2018,Chong2018}.
We define the perturbation parameter $0<\epsilon \ll 1$ and use the perturbative decomposition
\begin{subequations} \label{ansatz}
\begin{align}
    u_n&=\epsilon \left[A_{1,0}+\left(A_{1,1} E_n+{\rm c.c.}\right)\right]\nonumber\\
    &~~~+\epsilon^2\left[A_{2,0} + \left(A_{2,1} E_n+A_{2,2} E_n^2+{\rm c.c.}\right)\right]\nonumber\\
    &~~~+\epsilon ^3\left[A_{3,0}+\left(A_{3,1} E_n+A_{3,2} E_n^2+A_{3,3} E_n^3+{\rm c.c.}\right)\right],\\
    \varphi_n&=\epsilon\left[B_{1,0}+\left(B_{1,1} E_n+{\rm c.c.}\right)\right]\nonumber\\
    &~~~+\epsilon^2\left[B_{2,0}+\left(B_{2,1} E_n+B_{2,2} E_n^2+{\rm c.c.}\right)\right]\nonumber\\
    &~~~+\epsilon^3\left[B_{3,0}+\left(B_{3,1} E_n+B_{3,2} E_n^2+B_{3,3} E_n^3+{\rm c.c.}\right)\right],
\end{align}
\end{subequations}
where $E_n=E_n(t)=e^{i(kn-\omega t)}$, where $k$ and $\omega$ are the wave number and angular frequency, respectively 
and (c.c.) is the complex conjugate.
The $A_{i,j}=A_{i,j}(\xi,\tau)$ and $B_{i,j}=B_{i,j}(\xi,\tau)$ are amplitude functions to be determined that depend on the slow scale variables in space $\xi=\epsilon(n-\lambda t)$ and in time $\tau=\epsilon^2t$ with $\lambda$ being the group velocity.

Substituting ansatz \eqref{ansatz} into Eq.~\eqref{eq:eom_nondim} and collecting the terms according to the order of $\epsilon$ yields the wave dispersion relationship $\omega=\omega(k)$ at order $\mathcal{O}(\epsilon^1E_n^1)$,
\begin{align}
    \omega_{\pm}^2=2\left(\alpha_{11}+\alpha_{22}\pm\sqrt{\left(\alpha_{11}-\alpha_{22}\right)^2+(2\alpha_{12}\kappa)^2}\right)\nonumber\\
    \times\sin^2\left(\frac{k}{2}\right),\label{eq:dispersion}
\end{align}
where $\kappa^2=R_0^2/r^2=m_nR_0^2/j_n$ is the normalized curvature~(i.e., $r$ is a radius of gyration of the disc).

The wave dispersion relationship is shown in Fig.~{\ref{fig:schematic}}(b) for a few select sets of linear coefficients $\alpha_{11}$, $\alpha_{12}$, and $\alpha_{22}$.
If we keep the coupling term $\alpha_{12}=0$ and set $\alpha_{22} \neq \alpha_{11}$,
this results in two distinct curves denoted as red lines in Fig.~{\ref{fig:schematic}(b)}.
Similarly, if we let $\alpha_{11}\neq\alpha_{22}$, but now set $\alpha_{12} \neq 0$, the wave dispersion curves appears as two blue curves in Fig.~{\ref{fig:schematic}(b)}.

At the order $\mathcal{O}(\epsilon^2E_n^1)$, we obtain the group velocity $\lambda=d\omega/dk$,
\begin{align}
    \lambda=-\frac{1}{\omega_{\pm}}\left((\alpha_{11}+\alpha_{22})\pm\sqrt{\left(\alpha_{11}-\alpha_{22}\right)^2+(2\alpha_{12}\kappa)^2}\right)\nonumber\\\times\sin k.\label{eq:wavevelocity}
\end{align}
Finally, at order $\mathcal{O}(\epsilon^3E_n^1)$, nonlinear partial differential equations of $A_{1,1}$ and $B_{1,1}$ emerge,
\begin{widetext}
\begin{subequations}\label{eq:cnlse}
\begin{align}
    &i\partial_\tau A_{1,1}+\nu_2\partial_\xi^2 A_{1,1}+\nu_3\partial_\xi^2 B_{1,1}+\nu_4|A_{1,1}|^2A_{1,1}+\nu_5|B_{1,1}|^2B_{1,1}\nonumber\\
    &~~~~~~~~~~~~+\nu_6|B_{1,1}|^2A_{1,1}+\nu_7|A_{1,1}|^2B_{1,1}+\nu_8B_{1,1}^*A_{1,1}^2+\nu_9A_{1,1}^*B_{1,1}^2=0, \label{eq:nlse1}\\
    &i\partial_\tau B_{1,1}+\mu_2\partial_\xi^2 A_{1,1}+\mu_3\partial_\xi^2 B_{1,1}+\mu_4|A_{1,1}|^2A_{1,1}+\mu_5|B_{1,1}|^2B_{1,1}\nonumber\\
    &~~~~~~~~~~~~+\mu_6|B_{1,1}|^2A_{1,1}+\mu_7|A_{1,1}|^2B_{1,1}+\mu_8B_{1,1}^*A_{1,1}^2+\mu_9A_{1,1}^*B_{1,1}^2=0, \label{eq:nlse2}
\end{align}
\end{subequations}
\end{widetext}
where superscripts $(^*)$ denote the complex conjugate, and $\nu$ and $\mu$ with subscripts are the real constant coefficients defined in terms of the  coefficients $\alpha$~(see section 2 in the Supplementary Note for more details of the asymptotic expansion and section 3 therein for the detailed expressions of coefficients $\nu_i,~\mu_i$).
Note that Eq.~\eqref{eq:cnlse} resembles a coupled-NLSE, such as the Manakov system~\cite{Manakov1974}. Unlike the Manakov system, Eq.~\eqref{eq:cnlse} is non-integrable for generic values of the coefficients $\nu$ and $\mu$.

\section{Soliton Initial Data} \label{soliton}

\begin{figure*}[tbp]
    \centering
    \includegraphics[width=\linewidth]{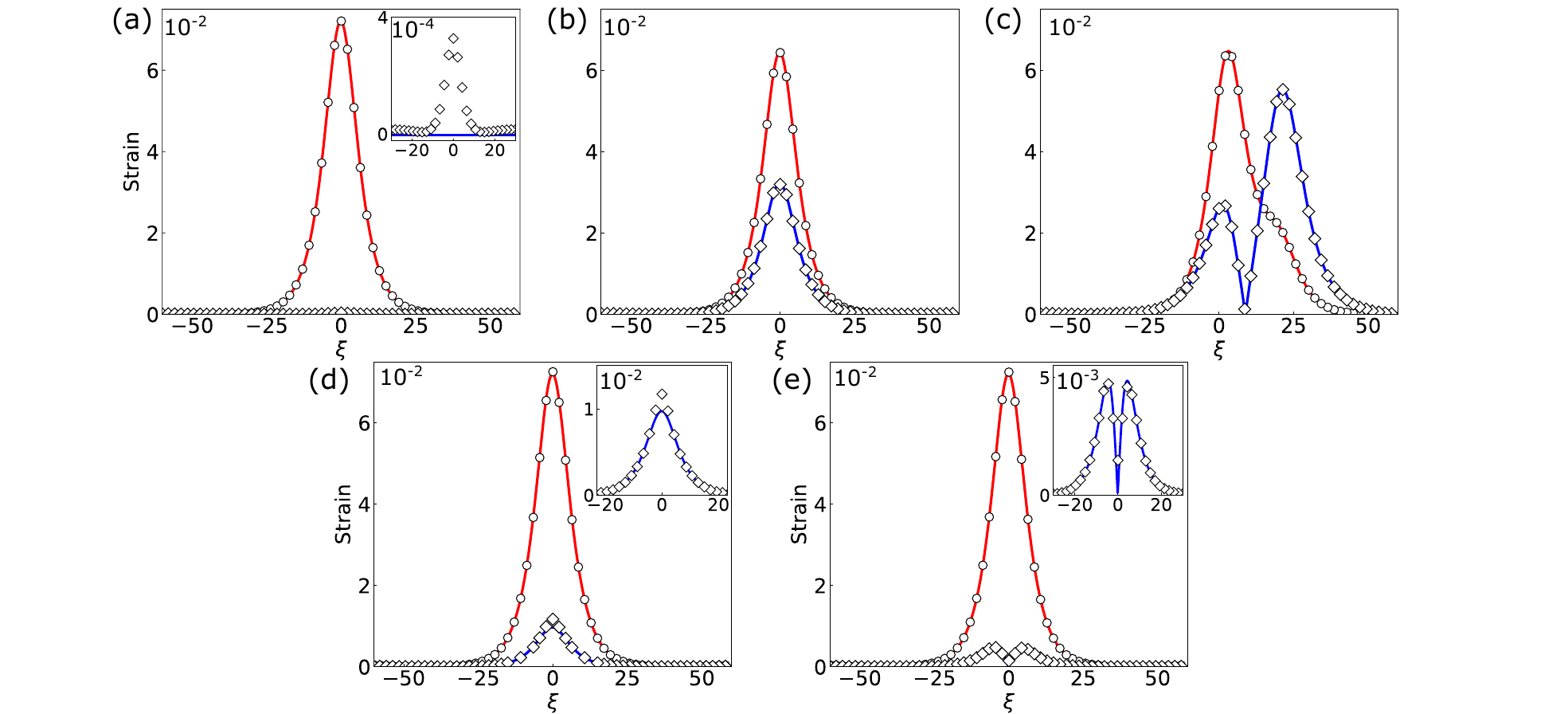}
    \caption{
    Coupled soliton solutions, which are analytically predicted (solid lines) and numerically computed for the full lattice equations and extracted at $\tau\approx8$ (open symbols). 
    (a) Initial data given by the single-component soliton of the
    two-component NLSE model, see Eq.~\eqref{eq:soliton_manakov1} with $P_1=1,~P_2=0$,
    (b) Initial data given by the two-component soliton with different amplitudes, see Eq.~\eqref{eq:soliton_manakov1} with $P_1=1,~P_2=0.5$,
    (c)  Initial data given by the incoherently-coupled NLSE soliton with bimodal $b$ component, see Eq.~\eqref{eq:soliton_manakov2},
    (d) Initial data given by the coherently-coupled NLSE soliton with unimodal $b$ component, see Eq.~\eqref{eq:soliton_coherent} with $m=0$, and 
    (e) Initial data given by the coherently-coupled NLSE soliton with bimodal $b$ component, see Eq.~\eqref{eq:soliton_coherent} with $m=1$.
    The parameters are $\epsilon=0.09$, $\epsilon_1=0.027$, $q=0.1$, $q_1=0.1$, and $q_2=0.08$.
    Red solid lines and open circles correspond to the axial mode; blue solid lines and open squares correspond to the rotational mode.
    The markers are plotted for every 50 spatial points. The insets are zooms of the rotational modes.}
    \label{fig:asoln_nsoln_soliton}
\end{figure*}

To start our investigation, we first consider two special cases where Eqs.~\eqref{eq:cnlse} reduce to well-known coupled NLSEs.
{In particular, we consider (i) the Manakov system and (ii) the coherently-coupled NLSE with energy exchange term.}
These special cases have exact solutions.

\subsection{Manakov Special Case}\label{manakov}
{If we let all NLSE coefficients be zero except for $\nu_2$, $\mu_3$, $\nu_4$, $\nu_6$, $\mu_7$, and $\mu_5$, Eq.~{\eqref{eq:cnlse}} reduces to the incoherently-coupled NLSE,}
\begin{subequations}\label{eq:manakov}
\begin{align}
    &i\partial_\tau A_{1,1}+\nu_2\partial_\xi^2A_{1,1}+\left(\nu_4|A_{1,1}|^2+\nu_6|B_{1,1}|^2\right)A_{1,1}=0, \label{eq:manakov1}\\
    &i\partial_\tau B_{1,1}+\mu_3\partial_\xi^2B_{1,1}+\left(\mu_7|A_{1,1}|^2+\mu_5|B_{1,1}|^2\right)B_{1,1}=0. \label{eq:manakov2}
\end{align}
\end{subequations}
The above equations are generally non-integrable except for a few special sets of coefficients~\mbox{\cite{Haelterman1993,Haelterman1994,Haelterman1994a}}. One of these is the well known Manakov system~\mbox{\cite{Manakov1974}}.
In this section, we consider the Manakov system with the coefficients $\nu_2=\mu_3=1/2$ and $\nu_4=\nu_6=\mu_7=\mu_5=1$, which has exact solutions of the form,
\begin{align}\label{eq:envelope1}
    \bcrm A_{1,1}(\xi,\tau)\\B_{1,1}(\xi,\tau)\ecrm=\bcrm a(\xi)\\b(\xi)\ecrm e^{i2q^2\tau},
\end{align}
where the envelopes $a$ and $b$ are real valued functions (without loss
of generality in the 1-dimensional case considered herein), and $q$ is a real parameter associated with the wave frequency.
{Among the many possible solutions of this form, we consider here the fundamental (bright) one-soliton solutions~\mbox{\cite{Manakov1974,Ablowitz2003}}},
\begin{align}\label{eq:soliton_manakov1}
    \bcrm a(\xi)\\b(\xi)\ecrm=\frac{2q}{\sqrt{P_1^2+P_2^2}}\bcrm P_1\\P_2\ecrm\sech(2q\xi).
\end{align}
Alternatively, if we assume
\begin{align}\label{eq:envelope2}
    \bcrm A_{1,1}(\xi,\tau)\\B_{1,1}(\xi,\tau)\ecrm=\bcrm a(\xi)e^{i2q_1^2\tau}\\b(\xi)e^{i2q_2^2\tau}\ecrm,
\end{align}
where two different real frequency parameters $q_1$ and $q_2$ exist, Eq.~\eqref{eq:manakov} allows the multi-hump soliton solutions~\mbox{\cite{Stalin2019,Ramakrishnan2020}},
\begin{align}\label{eq:soliton_manakov2}
    \bcrm a(\xi)\\b(\xi)\ecrm
        =&\frac{2}{F(\xi)}\left[
        \bcrm
        P_1e^{2q_1\xi}\\
        P_2e^{2q_2\xi}
        \ecrm\right.
        \nonumber\\
        &~~~~\left.+g
        \bcrm
         P_2q_1^2e^{2q_2\xi}\\
        -P_1q_2^2e^{2q_1\xi}
        \ecrm e^{2(q_1+q_2)\xi}\right]
\end{align}
where $P_1$ and $P_2$ are the arbitrary amplitude parameters, $F(\xi)=\frac{P_1^2}{4q_1^2}e^{4q_1\xi}+\frac{P_2^2}{4 q_2^2}e^{4q_2\xi}+\frac{P_1^2 P_2^2 \left(q_1-q_2\right)^2}{16q_1^2q_2^2\left(q_1+q_2\right)^2}e^{4\left(q_1+q_2\right)\xi}$, and $g=[P_1P_2\left(q_1-q_2\right)]/[(2q_1q_2)^2\left(q_1+q_2\right)]$.

\subsection{coherently-coupled NLSE System}

Another interesting example where solitary wave solutions can be identified is by setting the coefficients $\nu_3,~\nu_5,~\nu_7,~\nu_8$ and $\mu_2,~\mu_4,~\mu_6,~\mu_9$ of Eq.~\eqref{eq:cnlse} to zero.
Under such a selection,  Eq.~\eqref{eq:cnlse} reduces to the coherently-coupled NLSE~\cite{Kivshar2003} with the form:
\begin{subequations}\label{eq:ccnlse}
\begin{align}
    &i\partial_\tau A_{1,1}+\nu_2\partial_\xi^2A_{1,1}\nonumber\\
    &~~~+\left(\nu_4|A_{1,1}|^2+\nu_6|B_{1,1}|^2\right)A_{1,1}+\nu_9A_{1,1}^*B_{1,1}^2=0, \label{eq:ccnlse1}\\
    &i\partial_\tau B_{1,1}+\mu_3\partial_\xi^2B_{1,1}\nonumber\\
    &~~+\left(\mu_7|A_{1,1}|^2+\mu_5|B_{1,1}|^2\right)B_{1,1}+\mu_8B_{1,1}^*A_{1,1}^2=0. \label{eq:ccnlse2}
\end{align}
\end{subequations}
Once again, this is a model that frequently arises in nonlinear optics
in the realm of processes such as four-wave mixing and a systematic
derivation of such models can be found, e.g., in~\cite{Kivshar2003}.

When $\nu_2=\mu_3=1/2$ and $\nu_4=\mu_5=1$, this system also has solutions of the form given by Eq.~\eqref{eq:envelope1}~\mbox{\cite{Kivshar2003,Haelterman1993,Haelterman1994,Haelterman1994a}}, but now the amplitudes are only approximations,
\begin{subequations}\label{eq:soliton_coherent}
\begin{align}
    a(\xi)=&2q\sech(2q\xi),\label{eq:soliton_coherent_a}\\
    b(\xi)\approx&\epsilon_1\sqrt{1-G(\xi)^2}\pFq{2}{1}\left(-m,m+3,2,\frac{1-G(\xi)}{2}\right).\label{eq:soliton_coherent_b}
\end{align}
\end{subequations}

Here, $0<\epsilon_1\ll1$ is another perturbation parameter, $G(\xi)=\tanh(2q\xi)$, and $\pFq{2}{1}$ is a hypergeometric function.
In this expression, as is discussed in~\cite{Kivshar2003}, $m$ is a non-negative integer, and for each distinct corresponding value a different branch of vector solitons exists.
With the constraint that $m$ is an integer, in order for $m$-th order solitons to exist, the NLSE coefficients require the following relations, $\nu_9=(m+1)(m+2)-\nu_6,\quad \mu_9=(m+1)(m+2)-\mu_7$.

\subsection{Numerical simulations of coupled solitons} \label{numerics}
Figure~\ref{fig:asoln_nsoln_soliton} shows a comparison of analytical and numerical soliton solutions of axial and rotational components.
The lattice model initialized with various soliton solutions of the special cases considered above is numerically solved in the domain $\xi\in[-150,150]$ and $\tau\in[0,10]$ with perturbation parameter $\epsilon=0.09$.
Spatial profiles of the analytical and numerical solutions are extracted at $\tau=8$~(i.e., $t\approx987.7$), and plotted as solid lines and open symbols, respectively.
For the Manakov case, we used the coefficient values: $\alpha_{11}=\alpha_{22}=4$, $\alpha_{111}=-1$, $\alpha_{112}=\alpha_{122}=\alpha_{222}=\alpha_{1122}=\alpha_{1222}=1$, $\alpha_{1112}=\alpha_{2222}=2$, and $\alpha_{12}=\alpha_{1112}=0$.
For the coherently-coupled case, we used: for $n=0$, $\alpha_{11}=\alpha_{22}=4$, $\alpha_{11}=\alpha_{22}=3$, $\alpha_{112}=1$, $\alpha_{1111}=6$, $\alpha_{1112}=\alpha_{1122}=3/2$, $\alpha_{2222}=11/2$, and $\alpha_{12}=\alpha_{122}=\alpha_{1112}=0$.
For $n=1$, $\alpha_{11}=\alpha_{22}=4$, $\alpha_{111}=3$, $\alpha_{222}=19$, $\alpha_{112}=1$, $\alpha_{1111}=6$, $\alpha_{1112}=3/2$, $\alpha_{1122}=11/2$, $\alpha_{2222}=363/2$, and $\alpha_{12}=\alpha_{122}=\alpha_{1112}=0$.
In general, both numerically solved axial and rotational components agree well overall
with the analytical approximation, regardless of the initial condition or the choice of coefficients. There are however, also
deviations between the prediction and the actual dynamics, which is to be expected, given the approximate nature of the reduction. For example, there exists a small non-zero solution in the rotational component in Fig.~\ref{fig:asoln_nsoln_soliton}(a)~(see inset figure), despite initializing the lattice with a single-component solitary wave.
This suggests that there is a weak energy leakage from the axial channel (with non-zero initial data) to the rotational channel (with zero initial data).
However, given that the spatial profile in rotational mode is very small in amplitude, the relevant energy transfer is rather minimal.

When we have non-zero amplitude initial data in both axial and rotational components (Fig.~\ref{fig:asoln_nsoln_soliton}(b-e)), we see a good agreement with the analytical prediction, even for the case where either or both the axial and rotational component initial condition is asymmetric rather than unimodal~(Fig.~\ref{fig:asoln_nsoln_soliton}(c) and (e)).
There exist some slight disparities in the coherently-coupled NLSE case~(i.e., Fig.~\ref{fig:asoln_nsoln_soliton}(d-e)),
presumably due to the stronger coupling in coherently-coupled case.  Nevertheless, the overall agreement is excellent, regardless of the initial condition profile.

\section{Rogue Wave Initial Data} \label{peregrine}

Next, we consider solutions that are spatio-temporally localized, namely the rogue wave solutions of the two-component NLSE system (Eq.~{\eqref{eq:manakov}}).
Again, the NLSE coefficient $\nu_2=\mu_3=1/2$ and $\nu_4=\nu_6=\mu_7=\mu_5=1$ are chosen (i.e., the Manakov system).

One of the fundamental rogue wave solutions of the Manakov system~\cite{Baronio2012} is given by
\begin{align}\label{eq:crogue_manakov}
    \bcrm A_{1,1}(\xi,\tau)\\B_{1,1}(\xi,\tau)\ecrm=\left[L\bcrm P_1\\P_2\ecrm+M\bcrm P_2\\-P_1\ecrm\right]\frac{e^{i4q^2\tau}}{B},
\end{align}
where $a$ and $b$ are arbitrary real parameters, the real frequency parameter is $q=\sqrt{P_1^2+P_2^2}$, $L=\frac{3}{2}-32q^4\tau^2-8q^2\xi^2+i16q^2\tau+|f|^2e^{4q\xi}$, $M=4f(2q\xi-i4q^2\tau-\frac{1}{2})e^{2q\xi+i2q^2\tau}$, and $B=\frac{1}{2}+32q^4\tau^2+8q^2\xi^2+|f^2|e^{4q\xi}$ with $f$ being an arbitrary complex parameter.

\begin{figure*}[tbp]
    \centering
    \includegraphics[width=\linewidth]{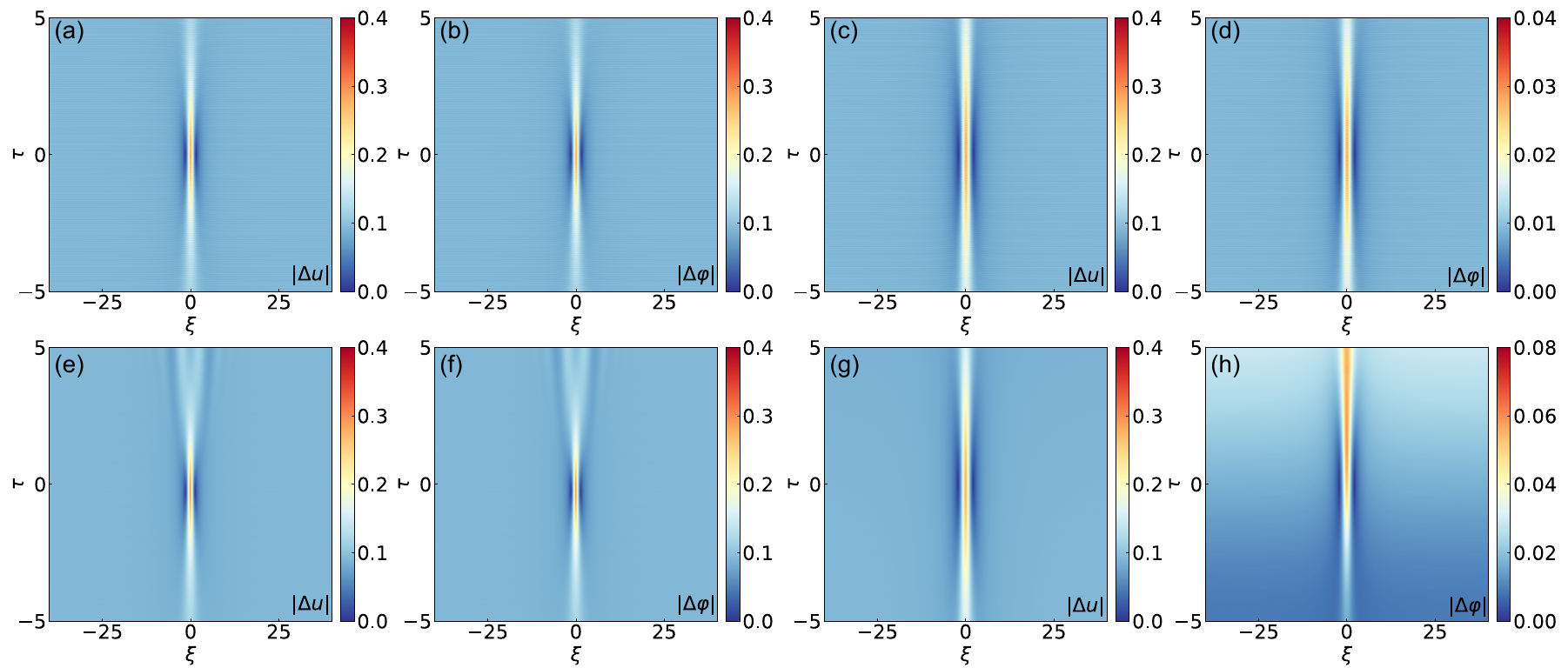}
    \caption{Coupled rogue wave solutions, which are analytically predicted (top row) and numerically solved for the full lattice equations initialized with Eq.~\eqref{eq:crogue_manakov} (bottom row).
    The perturbation parameter is $\epsilon=0.09$, and the amplitudes are (a,b,e,f) $P_1=P_2=0.2$, $f=0$, and 
    (c,d,g,h) $P_1=0.2$, $P_2=0.02$, and $f=0$.
    }
    \label{fig:asoln_nsoln_crogue}
\end{figure*}
Setting $f=0$, we obtain coupled vector solutions in the axial and rotational components that are reminiscent of the Peregrine soliton. We consider two 
case examples. One where the axial and rotational components are chosen to be identical (i.e., effectively the single component situation),
see Fig.~\ref{fig:asoln_nsoln_crogue}(a) and (b). We also consider a case example where the rotational component is $1/10$ of the amplitude of the axial component,
see Fig.~\ref{fig:asoln_nsoln_crogue}(c) and (d). In both cases, the peak amplitude is three times higher than the background and is localized at the origin. There are also density dips in the vicinity of the principal peak.

Using the spatial profile at $\tau=-5$ from the NLSE approximation as initial data, we simulate the lattice dynamics in the domain $\tau\in[-5,5]$ and $\xi\in[-40,40]$.
The perturbation parameter is set to $\epsilon=0.09$, and we choose the following coefficients: $\alpha_{11}=\alpha_{22}=4$, $\alpha_{122}=\sqrt{2}$, $\alpha_{1111}=\alpha_{1122}=1$, $\alpha_{2222}=2$, and $\alpha_{12}=\alpha_{111}=\alpha_{112}=\alpha_{222}=\alpha_{1112}=\alpha_{1222}=\alpha_{1222}=0$.
The resulting numerical solutions are shown in Fig.~\ref{fig:asoln_nsoln_crogue}(e-h).
In both components, the time until the localization coincides well with the analytical prediction, however there are slight discrepancies after the formation of the rogue wave (i.e., $\tau>0$).
In particular, the peak formed at the origin splits into smaller amplitude waves in the lattice case.
Similar observations have been made in other lattice settings~\cite{Charalampidis2018}. As discussed in Ref.~\cite{Charalampidis2018}, the formation of smaller waves may be induced by the modulational instability of the NLSE background, which is activated due to the large peak amplitude.

In Fig.~\ref{fig:asoln_nsoln_crogue}(g-h) where the axial and rotational component have different amplitudes, the waves tend to focus and thus localize at the origin. 
In the axial component, we see that the peak amplitude of the numerical solution is slightly lower than the analytical prediction.
On the contrary, the rotational component shows a peak that is twice as high as the analytical prediction.
The deviation from the analytical prediction suggests there is energy leakage from the axial component into the rotational component~(see Supplementary Note 4 for longer time spatio-temporal evolution and how it differs from analytical prediction).
We believe that this is due to the non-zero coupling terms of the lattice equation~(e.g., $\alpha_{122}$), which possibly trigger the energy transfer between two components, in a way that is not reflected in the reduced NLSE system.

\section{Gaussian Initial Data} \label{gauss}

To further explore rogue wave solutions in the coupled lattice, we hereafter numerically study Eq.~\eqref{eq:cnlse} in the more general case (i.e., with all coefficients being present).
However, as mentioned previously, Eq.~\eqref{eq:cnlse} is non-integrable, therefore no exact Peregrine-like solution
is analytically known. Thus, the lattice cannot be initialized with an analytical prediction to examine the time evolution.
As an alternative, we consider more general unimodal shaped data.
In particular, we use the Gaussian initialization, which has been shown to be effective in leading to rogue-like waves as a result of the gradient catastrophe phenomenon in the focusing NLSE~\cite{Bertola2013}.
This has been mathematically explored  originally in the so-called
semi-classical continuum NLSE system in the work~\mbox{\cite{Bertola2013}}, and more recently explored in 
corresponding experimental studies in nonlinear optics in
the work of~\cite{Tikan}.

Let the initial data be the Gaussian envelope function~\cite{Charalampidis2018},
\begin{align}
    \bcrm A_{1,1}(\xi,\tau=0)\\B_{1,1}(\xi,\tau=0)\ecrm
    =\bcrm P_1\\P_2\ecrm\exp\left(-\frac{\xi^2}{4\sigma^2}\right),
\end{align}
where $P_1$ and $P_2$ are arbitrary real parameters that determine the amplitude of the initial profile of $A_{1,1}$ and $B_{1,1}$ respectively, and $\sigma$ is the width of the localization.
The numerical simulation is then conducted in the domain $\tau\in[0,5]$ and $[0,20]$,
and $\xi\in[-30,30]$ with the perturbation parameter $\epsilon=0.09$.
The lattice coefficients are set to: $\alpha_{11}=\alpha_{22}=16$, $\alpha_{12}=0.016$, $\alpha_{111}=\alpha_{222}=\alpha_{1111}=\alpha_{1112}=\alpha_{1122}=\alpha_{1222}=\alpha_{2222}$, and $\alpha_{112}=\alpha_{122}=1.6$.
Here, these choices are made such that the NLSE becomes the focusing equation (i.e., $\nu_i>0$ and $\mu_i>0$).
The corresponding simulations of the NLSE~(Eq.~\eqref{eq:cnlse}) are also conducted as a reference solution to be compared with the lattice dynamics solutions.

\begin{figure}[tbp]
    \centering
    \includegraphics[width=\linewidth]{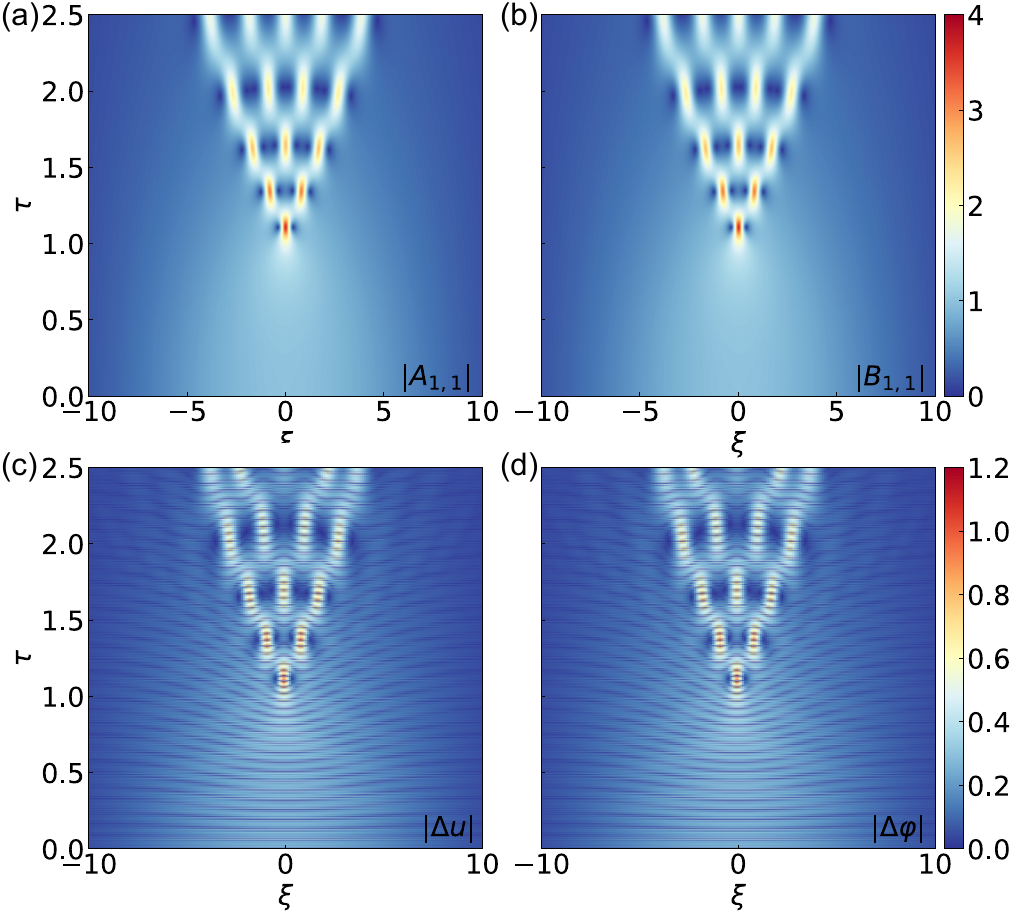}
    \caption{Numerical solutions of NLSE (top row) and of the lattice (bottom row) with all coefficients in Eq.~\eqref{eq:cnlse} being non-zero.
    The perturbation parameter is $\epsilon=0.09$, the width of the localization is $\sigma=4$, and the amplitudes are $P_1=P_2=1.0$.}
    \label{fig:nsoln_gaussian1}
\end{figure}
\begin{figure}[tbp]
    \centering
    \includegraphics[width=\linewidth]{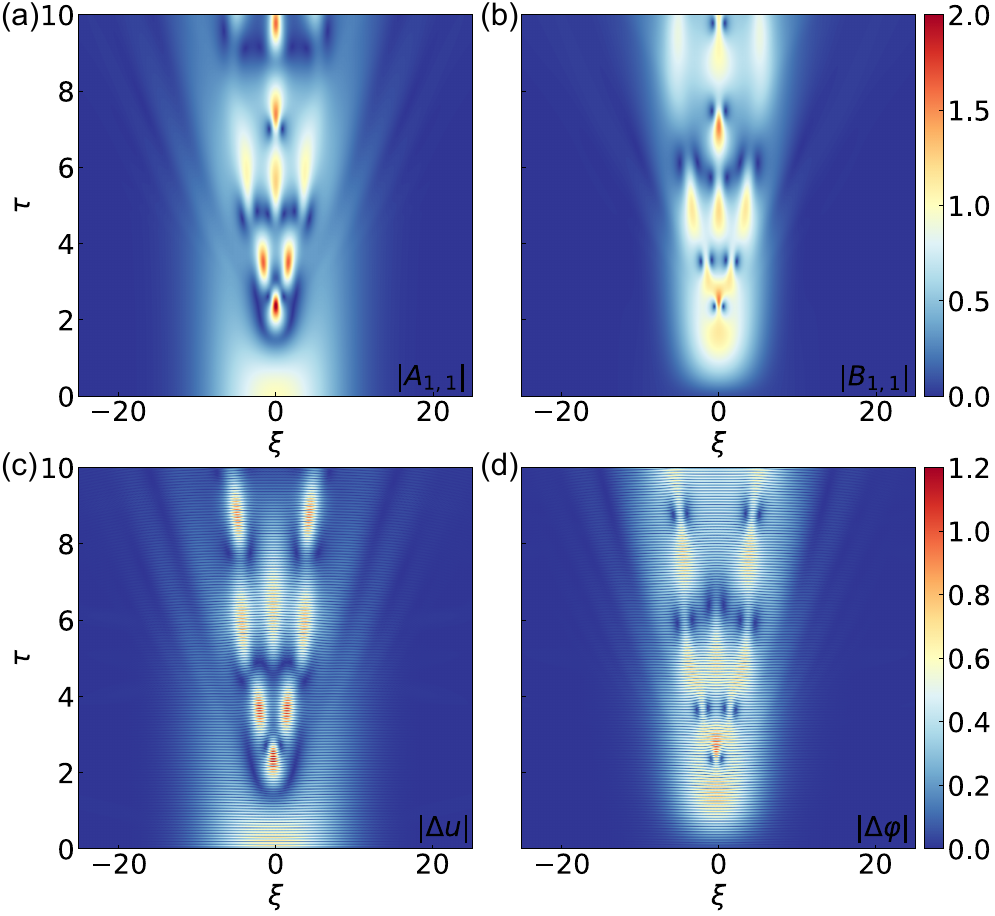}
    \caption{Numerical solutions of NLSE (a,b) and lattice (c,d) with all coefficients in Eq.~\eqref{eq:cnlse} being non-zero.
    Perturbation parameter is $\epsilon=0.2$, width of the localization is $\sigma=4$, and the amplitudes are $P_1=1.0,~P_2=0.1$.
    }
    \label{fig:nsoln_gaussian2}
\end{figure}

Figures~\ref{fig:nsoln_gaussian1} and \ref{fig:nsoln_gaussian2} show both lattice and NLSE simulation results for two different cases of initial conditions.
Specifically, (i) the initial condition where axial and rotational modes have equal amplitude (i.e., $P_1=P_2=1.0$; Fig.~\ref{fig:nsoln_gaussian1}), and (ii) the initial condition with the rotational mode being $1/10$ of axial mode (i.e., $P_1=1.0$, $P_2=0.1$; Fig.~\ref{fig:nsoln_gaussian2}) to examine how the energy transfer differs between the lattice and the NLSE simulation.
The localization width $\sigma=4$ is kept constant between the two cases.

First, if we use the equal amplitude initial conditions, the NLSE creates a tree-like pattern stemming from single peak localization at $\tau\approx1.1$.
This is in line with the integrable NLSE theory of~\cite{Bertola2013} and has also been observed in other systems, both continuum~\cite{Charalampidis2018a} and discrete~\cite{Charalampidis2018}.
The single peak localization has dips on the left and right side, which are directly reminiscent of a Peregrine soliton.
In the lattice simulation, we also see the tree-like pattern starting from the peak at $\tau\approx1.1$.
As can be observed, the branches formed after $\tau\approx2$ show small differences between the lattice simulation and the NLSE~(e.g., the two center peaks are formed slightly later in the lattice spatio-temporal evolution compared to the NLSE and the peak amplitude is different.
However, in general, the NLSE and lattice behave in a fairly similar manner, especially from the standpoint of the time at which the wave localizes in the early stage of the time evolution, the formation of the original Peregrine pattern, and also the tree-like pattern that follows.

Similarly, if we employ a smaller amplitude initial condition in the rotational mode, the NLSE and the lattice well agree in their spatio-temporal profiles.
In the NLSE simulation, the axial component~(Fig.~\ref{fig:nsoln_gaussian2}(a)), wide Gaussian initial data first decreases and then forms teardrop-like peak at $\tau\approx2.25$.
This single peak split into two peaks at $\tau\approx3.5$, then into three.
However, unlike the case with equal amplitude initial condition, the profile does not develop into tree-like pattern.
Instead, a more localized pattern centered around $\xi=0$ forms.
As for the rotational component of the NLSE shown in Fig.~\ref{fig:nsoln_gaussian2}(b), in constrast to the axial component, the amplitude first increases and reaches its highest amplitude at $\tau\approx1.5$ while keeping the broad width of the Gaussian initial profile.
Then, the peak narrows as two dips appear at $\tau\approx2.25$, where the axial component forms a single peak.
As time proceeds, we can observe two small and narrow peaks accompanied by adjacent dips that emerge at $\tau\approx3.5$.
In the lattice solution, we can observe similar dynamics.
For instance, the axial component forms a teardrop-like peak at $\tau\approx2.3$ while the rotational component shows two dips with a very narrow peak at $\xi=0$.
However, as we also observe in the equal amplitude initial data case, the spatio-temporal evolution of the lattice starts to deviate from the NLSE behavior as time proceeds.
After $\tau\approx6$ the pattern formations~(e.g., number of peaks) of the lattice deviate from those of NLSE.

\section{Energy exchange} \label{exchange}
\begin{figure}[tbp]
    \centering
    \includegraphics[width=\linewidth]{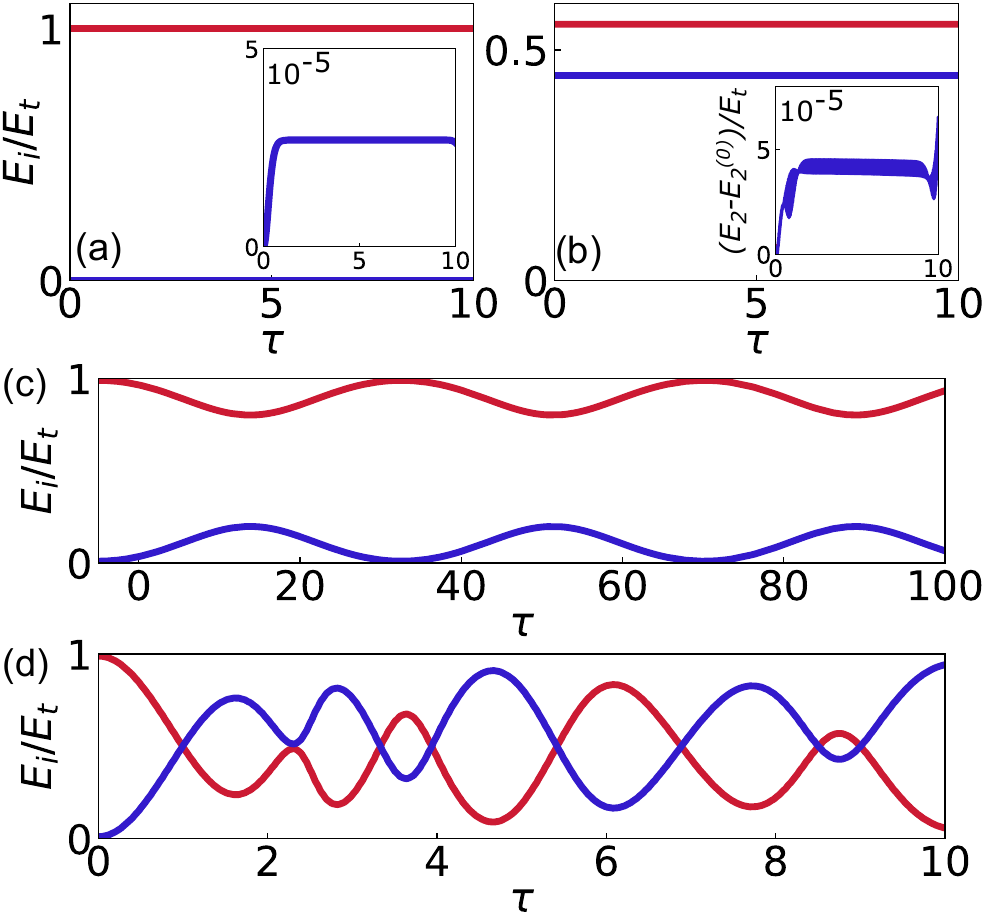}
    \caption{Energy profiles of axial $E_1$ and rotational $E_2$ component of the lattice, normalized by the total energy $E_t$.
    Red solid lines, axial component $E_1$; blue solid lines, rotational component $E_2$.
    Each panel corresponds to the soliton solutions: (a) Fig.~\ref{fig:asoln_nsoln_soliton}(a), (b): Fig.~\ref{fig:asoln_nsoln_soliton}(c), and rogue wave solutions (c): Fig.~\ref{fig:asoln_nsoln_crogue}(g-h), and Gaussian initial data solutions (d): Fig.~\ref{fig:nsoln_gaussian2}(c-d).
    The inset panel in: (a) represents a magnified view of the rotational mode in $\tau\in[0,10]$; (b) represents the deviation from the initial energy in rotational component, $[E_2-E_2(0)]/E_t$.
    }
    \label{fig:energy}
\end{figure}

In this section, we revisit the soliton and rogue wave solutions of the lattice simulation studied in the previous sections, and examine the energy profiles in the axial and rotational modes as a function of time.
We split the energy into two groups, (i) axial component and (ii) rotational components as follows:
\begin{subequations}
\begin{align}
    E_1&=\frac{1}{2}\dot{u}^2+\frac{1}{2}\alpha_{11}u^2+\frac{1}{6}\alpha_{111}u^3+\frac{1}{24}\alpha_{1111}u^4+\frac{1}{2}E_{cp},\\
    E_2&=\frac{1}{2}\dot{\varphi}^2+\frac{1}{2}\alpha_{22}\varphi^2+\frac{1}{6}\alpha_{222}\varphi^3+\frac{1}{24}\alpha_{2222}\varphi^4+\frac{1}{2}E_{cp},\\
    E_{cp}&=\alpha_{12}u\varphi+\frac{1}{2}\alpha_{112}u^2\varphi+\frac{1}{2}\alpha_{122}u\varphi^2\nonumber\\
        &~~~+\frac{1}{6}\alpha_{1112}u^3\varphi+\frac{1}{4}\alpha_{1122}u^2\varphi^2+\frac{1}{6}\alpha_{1222}u\varphi^3.
\end{align}
\end{subequations}
Note that we evenly distribute the energy due to coupling terms $E_{cp}$ among $E_1$ and $E_2$.
We investigate these two energy quantities for the solitary, rogue, and Gaussian induced wave solutions shown in Fig.~{\ref{fig:asoln_nsoln_soliton}}, Fig.~\ref{fig:asoln_nsoln_crogue}, and Fig.~{\ref{fig:nsoln_gaussian2}} respectively.
Figure~{\ref{fig:energy}} shows the energy of the axial and rotational component of different cases of the lattice simulation.

First, we take a closer look at the soliton solution case shown in Fig.~{\ref{fig:energy}}(a) and (b), which corresponds to the soliton solutions shown in Fig.~{\ref{fig:asoln_nsoln_soliton}(a)} and (c) respectively.
In general, both soliton solution energy profiles suggest that the energy does not transfer from one mode to another (i.e., $E_1$ and $E_2$ are constant throughout), except for the minimal leakage seen in the inset plot of Fig.~{\ref{fig:energy}}(a).
This energy leakage can also be seen in the spatial profile in Fig.~{\ref{fig:asoln_nsoln_soliton}}(a), where the rotational mode profile has a very small peak at the center.
The energy in the rotational mode rapidly increases from zero and then saturates, in this case around $E_2/E_t=2.5\times10^{-5}$.
Indeed, the stationary nature of the solution preserves the dynamics essentially thereafter.
Although we set the leading order coupling term $\alpha_{12}=0$, the lattice of interest is still coupled at higher orders (e.g., $\alpha_{112}u^2$ or $\alpha_{1112}u^3$).
Therefore, with non-zero axial amplitude $u$, the rotational mode is excited, and effectively the axial mode plays the role of an external potential of the small amplitude, leading to practically linear dynamics in the rotational mode.
Nonetheless, the energy leakage remains minimal in this case.
The profiles in Fig.~{\ref{fig:energy}(b)} also suggest the suppression of energy leakage since the energy profiles are almost constant, with minimal energy leakage from the axial to rotational the component.
In the inset panel of Fig.~\ref{fig:energy}(b), we see that the deviation from the initial value $\frac{E_2-E_2(0)}{E_t}$ is quite small~($\approx5\times10^{-5}$) and practically negligible.

In Fig.~\ref{fig:energy}(c) we show the time evolution of the energy component of the coupled rogue wave solution, which corresponds to the strain wave field in Fig.~\ref{fig:asoln_nsoln_crogue}(g,h), but with $\tau=35$.
We observe a continuous and gradual exchange of energy between the two channels.
As time progresses, the energy distributed to the rotational component grows and reaches its maximum, which is about 1/4 of the energy in axial component at $\tau\approx14$.
Then, $E_2$ decreases and attains a minimum at $\tau\approx32$.
Even in the longer term behavior, this gradual and partial exchange of the energy continues in a recurrent manner~(see also Supplementary Note~4 for the spatio-temporal evolution).

Lastly, we explore the energy exchange between the axial and
rotational components of the rogue wave-like solutions induced by Gaussian initial data, shown in Fig.~{\ref{fig:energy}}(d), which correspond to the strain wave field in Fig.~{\ref{fig:nsoln_gaussian2}}.
Unlike the above three cases, we see significant energy transfer between the two components.
As observed in the spatio-temporal evolution of the lattice solution, the axial and rotational component exchange a significant amount of energy quite quickly.
Indeed the rotational component of the energy $E_2$ overtakes the axial component at $\tau\approx1.6$.
When the first peak forms in the axial component of the lattice simulation~($\tau\approx2.3$; a narrow peak forms in rotational component), we see that the two energy components become almost identical.
Interestingly, even after the single peak formation, when the spatio-temporal profile of the lattice shows peaks, the difference between the two energy component becomes small.
For instance, two energy components essentially become identical again, when two peaks become significantly high in amplitude at $\tau\approx3.9$ in the axial component~(two narrow peaks form in the rotational component).
Similar behavior can also be observed at $\tau\approx6.9$ and $\tau\approx9$.

In summary, we observe three qualitatively different types of behavior of energy transfer.
For solitary wave initial data, there is minimal transfer of energy.
For Peregrine initial data, there is a partial transfer of energy between channels, and for Gaussian initial data, the energy is transferred continuously between the two channels in an aperiodic and oscillatory fashion.

\section{Conclusions and Future Work} \label{theend}
In conclusion, we have analytically and numerically explored nonlinear waves in a FPUT lattice with axial and rotational modes involving
up to cubic stiffness. We first derived coupled NLSE equations via a multiple-scale analysis. Variants of both incoherently-coupled and
coherently-coupled forms were considered and used to approximate the full lattice dynamics.
The approximation based on the solitary wave solution of the incoherently-coupled NLSE compared favorably to the numerical simulation of the  coupled FPUT lattice, both with and without energy exchange terms. In the coherently-coupled NLSE case, we also explored more complex waveforms in addition to the simplest unimodal solitary wave (where one component played the role of an effective potential for the other). Furthermore,  rogue wave type dynamics were studied.
First we used the exact coupled rogue wave solution of the incoherently-coupled NLSE system~(i.e., Manakov system), as the initial condition.
Regardless of the initial profile, the localization time of the analytical and numerical solution matched well, except for the small but noticeable energy leakage from the axial component to the rotational component.
When initialized with a sufficiently wide Gaussian envelope function, the lattice showed a clear localization due to the gradient catastrophe phenomenon, accompanied by the formation of secondary peaks, in line with a similar phenomenology previously analyzed in the NLSE realm.
Depending on the configuration and the initial data, coupled lattices of the FPUT type considered herein can effectively isolate the energy (e.g., soliton solutions) to one of the modes or continuously exchange the energy between modes while forming a peak~(e.g., Gaussian initial data solutions).

We believe that these findings open an analytical window of investigation of a multitude of systems that have recently been explored in various experiments at the multi-component setting~\cite{Yasuda2019,Deng2018,Deng2019}.
This allows one to observe  wave localization in a general coupled discrete nonlinear system, and may, in principle, open avenues to explore energy control in  mechanical systems.
At the same time, while here we presented the relevant multi-component technique at the one-dimensional, two-component setting, there are various recent works that suggest the relevance of corresponding considerations for higher numbers of components~\cite{Deng2019} or higher dimensions~\cite{Chong_2021}.

\section*{Acknowledgements}
The present paper is based on work that was supported by the US National Science Foundation under Grant Nos. 
CAREER-1553202 and CMMI-1933729 (JK), DMS-1809074 (PGK) and DMS-1615037 and DMS-2107945 (CC). Y.M. and J.Y. are grateful for the support of the Washington Research Foundation.

\bibliography{rogue_wave,tco}

\begin{thebibliography}{72}%
\makeatletter
\providecommand \@ifxundefined [1]{%
 \@ifx{#1\undefined}
}%
\providecommand \@ifnum [1]{%
 \ifnum #1\expandafter \@firstoftwo
 \else \expandafter \@secondoftwo
 \fi
}%
\providecommand \@ifx [1]{%
 \ifx #1\expandafter \@firstoftwo
 \else \expandafter \@secondoftwo
 \fi
}%
\providecommand \natexlab [1]{#1}%
\providecommand \enquote  [1]{``#1''}%
\providecommand \bibnamefont  [1]{#1}%
\providecommand \bibfnamefont [1]{#1}%
\providecommand \citenamefont [1]{#1}%
\providecommand \href@noop [0]{\@secondoftwo}%
\providecommand \href [0]{\begingroup \@sanitize@url \@href}%
\providecommand \@href[1]{\@@startlink{#1}\@@href}%
\providecommand \@@href[1]{\endgroup#1\@@endlink}%
\providecommand \@sanitize@url [0]{\catcode `\\12\catcode `\$12\catcode
  `\&12\catcode `\#12\catcode `\^12\catcode `\_12\catcode `\%12\relax}%
\providecommand \@@startlink[1]{}%
\providecommand \@@endlink[0]{}%
\providecommand \url  [0]{\begingroup\@sanitize@url \@url }%
\providecommand \@url [1]{\endgroup\@href {#1}{\urlprefix }}%
\providecommand \urlprefix  [0]{URL }%
\providecommand \Eprint [0]{\href }%
\providecommand \doibase [0]{http://dx.doi.org/}%
\providecommand \selectlanguage [0]{\@gobble}%
\providecommand \bibinfo  [0]{\@secondoftwo}%
\providecommand \bibfield  [0]{\@secondoftwo}%
\providecommand \translation [1]{[#1]}%
\providecommand \BibitemOpen [0]{}%
\providecommand \bibitemStop [0]{}%
\providecommand \bibitemNoStop [0]{.\EOS\space}%
\providecommand \EOS [0]{\spacefactor3000\relax}%
\providecommand \BibitemShut  [1]{\csname bibitem#1\endcsname}%
\let\auto@bib@innerbib\@empty
\bibitem [{\citenamefont {Kharif}\ and\ \citenamefont
  {Pelinovsky}(2003)}]{Kharif2003}%
  \BibitemOpen
  \bibfield  {author} {\bibinfo {author} {\bibfnamefont {C.}~\bibnamefont
  {Kharif}}\ and\ \bibinfo {author} {\bibfnamefont {E.}~\bibnamefont
  {Pelinovsky}},\ }\href {\doibase 10.1016/j.euromechflu.2003.09.002}
  {\bibfield  {journal} {\bibinfo  {journal} {European Journal of Mechanics,
  B/Fluids}\ }\textbf {\bibinfo {volume} {22}},\ \bibinfo {pages} {603}
  (\bibinfo {year} {2003})}\BibitemShut {NoStop}%
\bibitem [{\citenamefont {Haver}(2004)}]{Haver2004}%
  \BibitemOpen
  \bibfield  {author} {\bibinfo {author} {\bibfnamefont {S.}~\bibnamefont
  {Haver}},\ }\href
  {http://www.ifremer.fr/web-com/stw2004/rw/fullpapers/walk_on_haver.pdf}
  {\bibfield  {journal} {\bibinfo  {journal} {Rogue waves 2004 : proceedings of
  a workshop organized by Ifremer and held in Brest, France}\ }\textbf
  {\bibinfo {volume} {2004}},\ \bibinfo {pages} {1} (\bibinfo {year}
  {2004})}\BibitemShut {NoStop}%
\bibitem [{\citenamefont {Walker}\ \emph {et~al.}(2004)\citenamefont {Walker},
  \citenamefont {Taylor},\ and\ \citenamefont {Taylor}}]{Walker2004}%
  \BibitemOpen
  \bibfield  {author} {\bibinfo {author} {\bibfnamefont {D.~A.}\ \bibnamefont
  {Walker}}, \bibinfo {author} {\bibfnamefont {P.~H.}\ \bibnamefont {Taylor}},
  \ and\ \bibinfo {author} {\bibfnamefont {R.~E.}\ \bibnamefont {Taylor}},\
  }\href {\doibase 10.1016/j.apor.2005.02.001} {\bibfield  {journal} {\bibinfo
  {journal} {Applied Ocean Research}\ }\textbf {\bibinfo {volume} {26}},\
  \bibinfo {pages} {73} (\bibinfo {year} {2004})}\BibitemShut {NoStop}%
\bibitem [{\citenamefont {Adcock}\ \emph {et~al.}(2011)\citenamefont {Adcock},
  \citenamefont {Taylor}, \citenamefont {Yan}, \citenamefont {Ma},\ and\
  \citenamefont {Janssen}}]{Adcock2011}%
  \BibitemOpen
  \bibfield  {author} {\bibinfo {author} {\bibfnamefont {T.~A.}\ \bibnamefont
  {Adcock}}, \bibinfo {author} {\bibfnamefont {P.~H.}\ \bibnamefont {Taylor}},
  \bibinfo {author} {\bibfnamefont {S.}~\bibnamefont {Yan}}, \bibinfo {author}
  {\bibfnamefont {Q.~W.}\ \bibnamefont {Ma}}, \ and\ \bibinfo {author}
  {\bibfnamefont {P.~A.}\ \bibnamefont {Janssen}},\ }\href {\doibase
  10.1098/rspa.2011.0049} {\bibfield  {journal} {\bibinfo  {journal}
  {Proceedings of the Royal Society A: Mathematical, Physical and Engineering
  Sciences}\ }\textbf {\bibinfo {volume} {467}},\ \bibinfo {pages} {3004}
  (\bibinfo {year} {2011})}\BibitemShut {NoStop}%
\bibitem [{\citenamefont {Mori}\ and\ \citenamefont {Liu}(2002)}]{Mori2002}%
  \BibitemOpen
  \bibfield  {author} {\bibinfo {author} {\bibfnamefont {N.}~\bibnamefont
  {Mori}}\ and\ \bibinfo {author} {\bibfnamefont {P.~C.}\ \bibnamefont {Liu}},\
  }\href {\doibase 10.1016/S0029-8018(01)00073-7} {\bibfield  {journal}
  {\bibinfo  {journal} {Ocean Engineering}\ }\textbf {\bibinfo {volume} {29}},\
  \bibinfo {pages} {1399} (\bibinfo {year} {2002})}\BibitemShut {NoStop}%
\bibitem [{\citenamefont {Baschek}\ and\ \citenamefont
  {Imai}(2011)}]{Baschek2011}%
  \BibitemOpen
  \bibfield  {author} {\bibinfo {author} {\bibfnamefont {B.}~\bibnamefont
  {Baschek}}\ and\ \bibinfo {author} {\bibfnamefont {J.}~\bibnamefont {Imai}},\
  }\href {\doibase 10.5670/oceanog.2011.35} {\bibfield  {journal} {\bibinfo
  {journal} {Oceanography}\ }\textbf {\bibinfo {volume} {24}},\ \bibinfo
  {pages} {158} (\bibinfo {year} {2011})}\BibitemShut {NoStop}%
\bibitem [{\citenamefont {Pelinovsky}\ and\ \citenamefont
  {Kharif}(2016)}]{Pelinovsky2016}%
  \BibitemOpen
  \bibfield  {author} {\bibinfo {author} {\bibfnamefont {E.}~\bibnamefont
  {Pelinovsky}}\ and\ \bibinfo {author} {\bibfnamefont {C.}~\bibnamefont
  {Kharif}},\ }\href {\doibase 10.1007/978-3-319-21575-4} {\emph {\bibinfo
  {title} {Extreme Ocean Waves}}}\ (\bibinfo  {publisher} {Springer
  International Publishing},\ \bibinfo {address} {Cham},\ \bibinfo {year}
  {2016})\ pp.\ \bibinfo {pages} {1--236}\BibitemShut {NoStop}%
\bibitem [{\citenamefont {Chabchoub}\ \emph {et~al.}(2011)\citenamefont
  {Chabchoub}, \citenamefont {Hoffmann},\ and\ \citenamefont
  {Akhmediev}}]{Chabchoub2011}%
  \BibitemOpen
  \bibfield  {author} {\bibinfo {author} {\bibfnamefont {A.}~\bibnamefont
  {Chabchoub}}, \bibinfo {author} {\bibfnamefont {N.~P.}\ \bibnamefont
  {Hoffmann}}, \ and\ \bibinfo {author} {\bibfnamefont {N.}~\bibnamefont
  {Akhmediev}},\ }\href {\doibase 10.1103/PhysRevLett.106.204502} {\bibfield
  {journal} {\bibinfo  {journal} {Physical Review Letters}\ }\textbf {\bibinfo
  {volume} {106}},\ \bibinfo {pages} {204502} (\bibinfo {year}
  {2011})}\BibitemShut {NoStop}%
\bibitem [{\citenamefont {Chabchoub}\ \emph {et~al.}(2012)\citenamefont
  {Chabchoub}, \citenamefont {Hoffmann}, \citenamefont {Onorato},\ and\
  \citenamefont {Akhmediev}}]{Chabchoub2012}%
  \BibitemOpen
  \bibfield  {author} {\bibinfo {author} {\bibfnamefont {A.}~\bibnamefont
  {Chabchoub}}, \bibinfo {author} {\bibfnamefont {N.}~\bibnamefont {Hoffmann}},
  \bibinfo {author} {\bibfnamefont {M.}~\bibnamefont {Onorato}}, \ and\
  \bibinfo {author} {\bibfnamefont {N.}~\bibnamefont {Akhmediev}},\ }\href
  {\doibase 10.1103/PhysRevX.2.011015} {\bibfield  {journal} {\bibinfo
  {journal} {Physical Review X}\ }\textbf {\bibinfo {volume} {2}},\ \bibinfo
  {pages} {2} (\bibinfo {year} {2012})}\BibitemShut {NoStop}%
\bibitem [{\citenamefont {McAllister}\ \emph {et~al.}(2018)\citenamefont
  {McAllister}, \citenamefont {Draycott}, \citenamefont {Adcock}, \citenamefont
  {Taylor},\ and\ \citenamefont {{Van Den Bremer}}}]{Adcock2018}%
  \BibitemOpen
  \bibfield  {author} {\bibinfo {author} {\bibfnamefont {M.~L.}\ \bibnamefont
  {McAllister}}, \bibinfo {author} {\bibfnamefont {S.}~\bibnamefont
  {Draycott}}, \bibinfo {author} {\bibfnamefont {T.~A.}\ \bibnamefont
  {Adcock}}, \bibinfo {author} {\bibfnamefont {P.~H.}\ \bibnamefont {Taylor}},
  \ and\ \bibinfo {author} {\bibfnamefont {T.~S.}\ \bibnamefont {{Van Den
  Bremer}}},\ }\href {\doibase 10.1017/jfm.2018.886} {\bibfield  {journal}
  {\bibinfo  {journal} {Journal of Fluid Mechanics}\ }\textbf {\bibinfo
  {volume} {860}},\ \bibinfo {pages} {767} (\bibinfo {year}
  {2018})}\BibitemShut {NoStop}%
\bibitem [{\citenamefont {Xu}\ \emph {et~al.}(2020)\citenamefont {Xu},
  \citenamefont {Chabchoub}, \citenamefont {Pelinovsky},\ and\ \citenamefont
  {Kibler}}]{Chabchoub2020}%
  \BibitemOpen
  \bibfield  {author} {\bibinfo {author} {\bibfnamefont {G.}~\bibnamefont
  {Xu}}, \bibinfo {author} {\bibfnamefont {A.}~\bibnamefont {Chabchoub}},
  \bibinfo {author} {\bibfnamefont {D.~E.}\ \bibnamefont {Pelinovsky}}, \ and\
  \bibinfo {author} {\bibfnamefont {B.}~\bibnamefont {Kibler}},\ }\href
  {\doibase 10.1103/physrevresearch.2.033528} {\bibfield  {journal} {\bibinfo
  {journal} {Physical Review Research}\ }\textbf {\bibinfo {volume} {2}},\
  \bibinfo {pages} {33528} (\bibinfo {year} {2020})}\BibitemShut {NoStop}%
\bibitem [{\citenamefont {Charalampidis}\ \emph
  {et~al.}(2018{\natexlab{a}})\citenamefont {Charalampidis}, \citenamefont
  {Cuevas-Maraver}, \citenamefont {Frantzeskakis},\ and\ \citenamefont
  {Kevrekidis}}]{Charalampidis2018a}%
  \BibitemOpen
  \bibfield  {author} {\bibinfo {author} {\bibfnamefont {E.~G.}\ \bibnamefont
  {Charalampidis}}, \bibinfo {author} {\bibfnamefont {J.}~\bibnamefont
  {Cuevas-Maraver}}, \bibinfo {author} {\bibfnamefont {D.~J.}\ \bibnamefont
  {Frantzeskakis}}, \ and\ \bibinfo {author} {\bibfnamefont {P.~G.}\
  \bibnamefont {Kevrekidis}},\ }\href@noop {} {\bibfield  {journal} {\bibinfo
  {journal} {Romanian Reports in Physics}\ }\textbf {\bibinfo {volume} {70}},\
  \bibinfo {pages} {1} (\bibinfo {year} {2018}{\natexlab{a}})},\ \Eprint
  {http://arxiv.org/abs/1609.01798} {arXiv:1609.01798} \BibitemShut {NoStop}%
\bibitem [{\citenamefont {Solli}\ \emph {et~al.}(2007)\citenamefont {Solli},
  \citenamefont {Ropers}, \citenamefont {Koonath},\ and\ \citenamefont
  {Jalali}}]{Solli2007}%
  \BibitemOpen
  \bibfield  {author} {\bibinfo {author} {\bibfnamefont {D.~R.}\ \bibnamefont
  {Solli}}, \bibinfo {author} {\bibfnamefont {C.}~\bibnamefont {Ropers}},
  \bibinfo {author} {\bibfnamefont {P.}~\bibnamefont {Koonath}}, \ and\
  \bibinfo {author} {\bibfnamefont {B.}~\bibnamefont {Jalali}},\ }\href
  {\doibase 10.1038/nature06402} {\bibfield  {journal} {\bibinfo  {journal}
  {Nature}\ }\textbf {\bibinfo {volume} {450}},\ \bibinfo {pages} {1054}
  (\bibinfo {year} {2007})}\BibitemShut {NoStop}%
\bibitem [{\citenamefont {Dudley}\ \emph {et~al.}(2014)\citenamefont {Dudley},
  \citenamefont {Dias}, \citenamefont {Erkintalo},\ and\ \citenamefont
  {Genty}}]{dudley1}%
  \BibitemOpen
  \bibfield  {author} {\bibinfo {author} {\bibfnamefont {J.~M.}\ \bibnamefont
  {Dudley}}, \bibinfo {author} {\bibfnamefont {F.}~\bibnamefont {Dias}},
  \bibinfo {author} {\bibfnamefont {M.}~\bibnamefont {Erkintalo}}, \ and\
  \bibinfo {author} {\bibfnamefont {G.}~\bibnamefont {Genty}},\ }\href
  {\doibase 10.1038/nphoton.2014.220} {\bibfield  {journal} {\bibinfo
  {journal} {Nature Photonics}\ }\textbf {\bibinfo {volume} {8}},\ \bibinfo
  {pages} {755} (\bibinfo {year} {2014})}\BibitemShut {NoStop}%
\bibitem [{\citenamefont {Frisquet}\ \emph {et~al.}(2016)\citenamefont
  {Frisquet}, \citenamefont {Kibler}, \citenamefont {Morin}, \citenamefont
  {Baronio}, \citenamefont {Conforti}, \citenamefont {Millot},\ and\
  \citenamefont {Wabnitz}}]{Kibler2016}%
  \BibitemOpen
  \bibfield  {author} {\bibinfo {author} {\bibfnamefont {B.}~\bibnamefont
  {Frisquet}}, \bibinfo {author} {\bibfnamefont {B.}~\bibnamefont {Kibler}},
  \bibinfo {author} {\bibfnamefont {P.}~\bibnamefont {Morin}}, \bibinfo
  {author} {\bibfnamefont {F.}~\bibnamefont {Baronio}}, \bibinfo {author}
  {\bibfnamefont {M.}~\bibnamefont {Conforti}}, \bibinfo {author}
  {\bibfnamefont {G.}~\bibnamefont {Millot}}, \ and\ \bibinfo {author}
  {\bibfnamefont {S.}~\bibnamefont {Wabnitz}},\ }\href {\doibase
  10.1038/srep20785} {\bibfield  {journal} {\bibinfo  {journal} {Scientific
  Reports}\ }\textbf {\bibinfo {volume} {6}},\ \bibinfo {pages} {1} (\bibinfo
  {year} {2016})}\BibitemShut {NoStop}%
\bibitem [{\citenamefont {Tikan}\ \emph {et~al.}(2017)\citenamefont {Tikan},
  \citenamefont {Billet}, \citenamefont {El}, \citenamefont {Tovbis},
  \citenamefont {Bertola}, \citenamefont {Sylvestre}, \citenamefont {Gustave},
  \citenamefont {Randoux}, \citenamefont {Genty}, \citenamefont {Suret},\ and\
  \citenamefont {Dudley}}]{Tikan}%
  \BibitemOpen
  \bibfield  {author} {\bibinfo {author} {\bibfnamefont {A.}~\bibnamefont
  {Tikan}}, \bibinfo {author} {\bibfnamefont {C.}~\bibnamefont {Billet}},
  \bibinfo {author} {\bibfnamefont {G.}~\bibnamefont {El}}, \bibinfo {author}
  {\bibfnamefont {A.}~\bibnamefont {Tovbis}}, \bibinfo {author} {\bibfnamefont
  {M.}~\bibnamefont {Bertola}}, \bibinfo {author} {\bibfnamefont
  {T.}~\bibnamefont {Sylvestre}}, \bibinfo {author} {\bibfnamefont
  {F.}~\bibnamefont {Gustave}}, \bibinfo {author} {\bibfnamefont
  {S.}~\bibnamefont {Randoux}}, \bibinfo {author} {\bibfnamefont
  {G.}~\bibnamefont {Genty}}, \bibinfo {author} {\bibfnamefont
  {P.}~\bibnamefont {Suret}}, \ and\ \bibinfo {author} {\bibfnamefont {J.~M.}\
  \bibnamefont {Dudley}},\ }\href {\doibase 10.1103/PhysRevLett.119.033901}
  {\bibfield  {journal} {\bibinfo  {journal} {Physical Review Letters}\
  }\textbf {\bibinfo {volume} {119}},\ \bibinfo {pages} {33901} (\bibinfo
  {year} {2017})}\BibitemShut {NoStop}%
\bibitem [{\citenamefont {H{\"{o}}hmann}\ \emph {et~al.}(2010)\citenamefont
  {H{\"{o}}hmann}, \citenamefont {Kuhl}, \citenamefont {St{\"{o}}ckmann},
  \citenamefont {Kaplan},\ and\ \citenamefont {Heller}}]{Hohmann2010}%
  \BibitemOpen
  \bibfield  {author} {\bibinfo {author} {\bibfnamefont {R.}~\bibnamefont
  {H{\"{o}}hmann}}, \bibinfo {author} {\bibfnamefont {U.}~\bibnamefont {Kuhl}},
  \bibinfo {author} {\bibfnamefont {H.~J.}\ \bibnamefont {St{\"{o}}ckmann}},
  \bibinfo {author} {\bibfnamefont {L.}~\bibnamefont {Kaplan}}, \ and\ \bibinfo
  {author} {\bibfnamefont {E.~J.}\ \bibnamefont {Heller}},\ }\href {\doibase
  10.1103/PhysRevLett.104.093901} {\bibfield  {journal} {\bibinfo  {journal}
  {Physical Review Letters}\ }\textbf {\bibinfo {volume} {104}},\ \bibinfo
  {pages} {1} (\bibinfo {year} {2010})},\ \Eprint
  {http://arxiv.org/abs/0909.0847} {arXiv:0909.0847} \BibitemShut {NoStop}%
\bibitem [{\citenamefont {Ruderman}(2010)}]{Ruderman2010}%
  \BibitemOpen
  \bibfield  {author} {\bibinfo {author} {\bibfnamefont {M.~S.}\ \bibnamefont
  {Ruderman}},\ }\href {\doibase 10.1140/epjst/e2010-01238-7} {\bibfield
  {journal} {\bibinfo  {journal} {European Physical Journal: Special Topics}\
  }\textbf {\bibinfo {volume} {185}},\ \bibinfo {pages} {57} (\bibinfo {year}
  {2010})}\BibitemShut {NoStop}%
\bibitem [{\citenamefont {Sabry}\ \emph {et~al.}(2012)\citenamefont {Sabry},
  \citenamefont {Moslem},\ and\ \citenamefont {Shukla}}]{Sabry2012}%
  \BibitemOpen
  \bibfield  {author} {\bibinfo {author} {\bibfnamefont {R.}~\bibnamefont
  {Sabry}}, \bibinfo {author} {\bibfnamefont {W.~M.}\ \bibnamefont {Moslem}}, \
  and\ \bibinfo {author} {\bibfnamefont {P.~K.}\ \bibnamefont {Shukla}},\
  }\href {\doibase 10.1063/1.4772058} {\bibfield  {journal} {\bibinfo
  {journal} {Physics of Plasmas}\ }\textbf {\bibinfo {volume} {19}} (\bibinfo
  {year} {2012}),\ 10.1063/1.4772058}\BibitemShut {NoStop}%
\bibitem [{\citenamefont {Bains}\ \emph {et~al.}(2014)\citenamefont {Bains},
  \citenamefont {Li},\ and\ \citenamefont {Xia}}]{Bains2014}%
  \BibitemOpen
  \bibfield  {author} {\bibinfo {author} {\bibfnamefont {A.~S.}\ \bibnamefont
  {Bains}}, \bibinfo {author} {\bibfnamefont {B.}~\bibnamefont {Li}}, \ and\
  \bibinfo {author} {\bibfnamefont {L.~D.}\ \bibnamefont {Xia}},\ }\href
  {\doibase 10.1063/1.4869464} {\bibfield  {journal} {\bibinfo  {journal}
  {Physics of Plasmas}\ }\textbf {\bibinfo {volume} {21}} (\bibinfo {year}
  {2014}),\ 10.1063/1.4869464},\ \Eprint {http://arxiv.org/abs/1403.3745}
  {arXiv:1403.3745} \BibitemShut {NoStop}%
\bibitem [{\citenamefont {Tolba}\ \emph {et~al.}(2015)\citenamefont {Tolba},
  \citenamefont {Moslem}, \citenamefont {El-Bedwehy},\ and\ \citenamefont
  {El-Labany}}]{Tolba2015}%
  \BibitemOpen
  \bibfield  {author} {\bibinfo {author} {\bibfnamefont {R.~E.}\ \bibnamefont
  {Tolba}}, \bibinfo {author} {\bibfnamefont {W.~M.}\ \bibnamefont {Moslem}},
  \bibinfo {author} {\bibfnamefont {N.~A.}\ \bibnamefont {El-Bedwehy}}, \ and\
  \bibinfo {author} {\bibfnamefont {S.~K.}\ \bibnamefont {El-Labany}},\ }\href
  {\doibase 10.1063/1.4918706} {\bibfield  {journal} {\bibinfo  {journal}
  {Physics of Plasmas}\ }\textbf {\bibinfo {volume} {22}} (\bibinfo {year}
  {2015}),\ 10.1063/1.4918706}\BibitemShut {NoStop}%
\bibitem [{\citenamefont {Onorato}\ \emph {et~al.}(2013)\citenamefont
  {Onorato}, \citenamefont {Residori}, \citenamefont {Bortolozzo},
  \citenamefont {Montina},\ and\ \citenamefont {Arecchi}}]{Onorato2013}%
  \BibitemOpen
  \bibfield  {author} {\bibinfo {author} {\bibfnamefont {M.}~\bibnamefont
  {Onorato}}, \bibinfo {author} {\bibfnamefont {S.}~\bibnamefont {Residori}},
  \bibinfo {author} {\bibfnamefont {U.}~\bibnamefont {Bortolozzo}}, \bibinfo
  {author} {\bibfnamefont {A.}~\bibnamefont {Montina}}, \ and\ \bibinfo
  {author} {\bibfnamefont {F.~T.}\ \bibnamefont {Arecchi}},\ }\href {\doibase
  10.1016/j.physrep.2013.03.001} {\enquote {\bibinfo {title} {{Rogue waves and
  their generating mechanisms in different physical contexts}},}\ } (\bibinfo
  {year} {2013})\BibitemShut {NoStop}%
\bibitem [{\citenamefont {{R. Osborne}}(2010)}]{R.Osborne2010}%
  \BibitemOpen
  \bibfield  {author} {\bibinfo {author} {\bibfnamefont {A.}~\bibnamefont {{R.
  Osborne}}},\ }\href {\doibase 10.1016/b978-012613760-6/50033-4} {\emph
  {\bibinfo {title} {Nonlinear Ocean Wave and the Inverse Scattering
  Transform}}}\ (\bibinfo  {publisher} {Elsevier},\ \bibinfo {address}
  {Amsterdam},\ \bibinfo {year} {2010})\BibitemShut {NoStop}%
\bibitem [{\citenamefont {Sulem}\ and\ \citenamefont
  {Sulem}(1999)}]{Sulem1999}%
  \BibitemOpen
  \bibfield  {author} {\bibinfo {author} {\bibfnamefont {C.}~\bibnamefont
  {Sulem}}\ and\ \bibinfo {author} {\bibfnamefont {P.}~\bibnamefont {Sulem}},\
  }\href {\doibase 10.1007/b98958} {\emph {\bibinfo {title} {The Nonlinear
  Schr{\"{o}}dinger Equation: Self-Focusing and Wave Collapse}}}\ (\bibinfo
  {publisher} {Springer-Verlag New York},\ \bibinfo {year} {1999})\BibitemShut
  {NoStop}%
\bibitem [{\citenamefont {Ablowitz}\ \emph {et~al.}(2003)\citenamefont
  {Ablowitz}, \citenamefont {Prinari},\ and\ \citenamefont
  {Trubatch}}]{Ablowitz2003}%
  \BibitemOpen
  \bibfield  {author} {\bibinfo {author} {\bibfnamefont {M.~J.}\ \bibnamefont
  {Ablowitz}}, \bibinfo {author} {\bibfnamefont {B.}~\bibnamefont {Prinari}}, \
  and\ \bibinfo {author} {\bibfnamefont {A.~D.}\ \bibnamefont {Trubatch}},\
  }\href {\doibase 10.1017/cbo9780511546709} {\emph {\bibinfo {title} {Discrete
  and Continuous Nonlinear Schr{\"{o}}dinger Systems}}}\ (\bibinfo  {publisher}
  {Cambridge University Press},\ \bibinfo {year} {2003})\BibitemShut {NoStop}%
\bibitem [{\citenamefont {Peregrine}(1983)}]{Peregrine1983}%
  \BibitemOpen
  \bibfield  {author} {\bibinfo {author} {\bibfnamefont {D.~H.}\ \bibnamefont
  {Peregrine}},\ }\href {\doibase 10.1017/s0334270000003891} {\bibfield
  {journal} {\bibinfo  {journal} {The Journal of the Australian Mathematical
  Society. Series B. Applied Mathematics}\ }\textbf {\bibinfo {volume} {25}},\
  \bibinfo {pages} {16} (\bibinfo {year} {1983})}\BibitemShut {NoStop}%
\bibitem [{\citenamefont {Charalampidis}\ \emph
  {et~al.}(2018{\natexlab{b}})\citenamefont {Charalampidis}, \citenamefont
  {Lee}, \citenamefont {Kevrekidis},\ and\ \citenamefont
  {Chong}}]{Charalampidis2018}%
  \BibitemOpen
  \bibfield  {author} {\bibinfo {author} {\bibfnamefont {E.~G.}\ \bibnamefont
  {Charalampidis}}, \bibinfo {author} {\bibfnamefont {J.}~\bibnamefont {Lee}},
  \bibinfo {author} {\bibfnamefont {P.~G.}\ \bibnamefont {Kevrekidis}}, \ and\
  \bibinfo {author} {\bibfnamefont {C.}~\bibnamefont {Chong}},\ }\href
  {\doibase 10.1103/PhysRevE.98.032903} {\bibfield  {journal} {\bibinfo
  {journal} {Physical Review E}\ }\textbf {\bibinfo {volume} {98}},\ \bibinfo
  {pages} {1} (\bibinfo {year} {2018}{\natexlab{b}})},\ \Eprint
  {http://arxiv.org/abs/1801.06086} {arXiv:1801.06086} \BibitemShut {NoStop}%
\bibitem [{\citenamefont {Akhmediev}\ and\ \citenamefont
  {Ankiewicz}(2011)}]{Akhmediev2011}%
  \BibitemOpen
  \bibfield  {author} {\bibinfo {author} {\bibfnamefont {N.}~\bibnamefont
  {Akhmediev}}\ and\ \bibinfo {author} {\bibfnamefont {A.}~\bibnamefont
  {Ankiewicz}},\ }\href {\doibase 10.1103/PhysRevE.83.046603} {\bibfield
  {journal} {\bibinfo  {journal} {Physical Review E - Statistical, Nonlinear,
  and Soft Matter Physics}\ }\textbf {\bibinfo {volume} {83}},\ \bibinfo
  {pages} {1} (\bibinfo {year} {2011})}\BibitemShut {NoStop}%
\bibitem [{\citenamefont {Ankiewicz}\ \emph {et~al.}(2010)\citenamefont
  {Ankiewicz}, \citenamefont {Akhmediev},\ and\ \citenamefont
  {Soto-Crespo}}]{Ankiewicz2010}%
  \BibitemOpen
  \bibfield  {author} {\bibinfo {author} {\bibfnamefont {A.}~\bibnamefont
  {Ankiewicz}}, \bibinfo {author} {\bibfnamefont {N.}~\bibnamefont
  {Akhmediev}}, \ and\ \bibinfo {author} {\bibfnamefont {J.~M.}\ \bibnamefont
  {Soto-Crespo}},\ }\href {\doibase 10.1103/PhysRevE.82.026602} {\bibfield
  {journal} {\bibinfo  {journal} {Physical Review E - Statistical, Nonlinear,
  and Soft Matter Physics}\ }\textbf {\bibinfo {volume} {82}} (\bibinfo {year}
  {2010}),\ 10.1103/PhysRevE.82.026602}\BibitemShut {NoStop}%
\bibitem [{\citenamefont {Wen}\ and\ \citenamefont {Wang}(2018)}]{Wen2018}%
  \BibitemOpen
  \bibfield  {author} {\bibinfo {author} {\bibfnamefont {X.~Y.}\ \bibnamefont
  {Wen}}\ and\ \bibinfo {author} {\bibfnamefont {D.~S.}\ \bibnamefont {Wang}},\
  }\href {\doibase 10.1016/j.wavemoti.2018.03.004} {\bibfield  {journal}
  {\bibinfo  {journal} {Wave Motion}\ }\textbf {\bibinfo {volume} {79}},\
  \bibinfo {pages} {84} (\bibinfo {year} {2018})}\BibitemShut {NoStop}%
\bibitem [{\citenamefont {Yan}\ and\ \citenamefont {Jiang}(2012)}]{Yan2012}%
  \BibitemOpen
  \bibfield  {author} {\bibinfo {author} {\bibfnamefont {Z.}~\bibnamefont
  {Yan}}\ and\ \bibinfo {author} {\bibfnamefont {D.}~\bibnamefont {Jiang}},\
  }\href {\doibase 10.1016/j.jmaa.2012.05.058} {\bibfield  {journal} {\bibinfo
  {journal} {Journal of Mathematical Analysis and Applications}\ }\textbf
  {\bibinfo {volume} {395}},\ \bibinfo {pages} {542} (\bibinfo {year}
  {2012})}\BibitemShut {NoStop}%
\bibitem [{\citenamefont {Maluckov}\ \emph {et~al.}(2013)\citenamefont
  {Maluckov}, \citenamefont {Lazarides}, \citenamefont {Tsironis},\ and\
  \citenamefont {Had{\v{z}}ievski}}]{Maluckov2013}%
  \BibitemOpen
  \bibfield  {author} {\bibinfo {author} {\bibfnamefont {A.}~\bibnamefont
  {Maluckov}}, \bibinfo {author} {\bibfnamefont {N.}~\bibnamefont {Lazarides}},
  \bibinfo {author} {\bibfnamefont {G.~P.}\ \bibnamefont {Tsironis}}, \ and\
  \bibinfo {author} {\bibfnamefont {L.}~\bibnamefont {Had{\v{z}}ievski}},\
  }\href {\doibase 10.1016/j.physd.2013.03.001} {\bibfield  {journal} {\bibinfo
   {journal} {Physica D: Nonlinear Phenomena}\ }\textbf {\bibinfo {volume}
  {252}},\ \bibinfo {pages} {59} (\bibinfo {year} {2013})},\ \Eprint
  {http://arxiv.org/abs/1204.5303} {arXiv:1204.5303} \BibitemShut {NoStop}%
\bibitem [{\citenamefont {Hoffmann}\ \emph {et~al.}(2018)\citenamefont
  {Hoffmann}, \citenamefont {Charalampidis}, \citenamefont {Frantzeskakis},\
  and\ \citenamefont {Kevrekidis}}]{HOFFMANN20183064}%
  \BibitemOpen
  \bibfield  {author} {\bibinfo {author} {\bibfnamefont {C.}~\bibnamefont
  {Hoffmann}}, \bibinfo {author} {\bibfnamefont {E.~G.}\ \bibnamefont
  {Charalampidis}}, \bibinfo {author} {\bibfnamefont {D.~J.}\ \bibnamefont
  {Frantzeskakis}}, \ and\ \bibinfo {author} {\bibfnamefont {P.~G.}\
  \bibnamefont {Kevrekidis}},\ }\href {\doibase 10.1016/j.physleta.2018.08.014}
  {\bibfield  {journal} {\bibinfo  {journal} {Physics Letters, Section A:
  General, Atomic and Solid State Physics}\ }\textbf {\bibinfo {volume}
  {382}},\ \bibinfo {pages} {3064} (\bibinfo {year} {2018})}\BibitemShut
  {NoStop}%
\bibitem [{\citenamefont {Sullivan}\ \emph {et~al.}(2020)\citenamefont
  {Sullivan}, \citenamefont {Charalampidis}, \citenamefont {Cuevas-Maraver},
  \citenamefont {Kevrekidis},\ and\ \citenamefont {Karachalios}}]{sullivan}%
  \BibitemOpen
  \bibfield  {author} {\bibinfo {author} {\bibfnamefont {J.}~\bibnamefont
  {Sullivan}}, \bibinfo {author} {\bibfnamefont {E.}~\bibnamefont
  {Charalampidis}}, \bibinfo {author} {\bibfnamefont {J.}~\bibnamefont
  {Cuevas-Maraver}}, \bibinfo {author} {\bibfnamefont {P.}~\bibnamefont
  {Kevrekidis}}, \ and\ \bibinfo {author} {\bibfnamefont {N.}~\bibnamefont
  {Karachalios}},\ }\href {\doibase 10.1140/epjp/s13360-020-00596-1} {\bibfield
   {journal} {\bibinfo  {journal} {European Physical Journal Plus}\ }\textbf
  {\bibinfo {volume} {135}},\ \bibinfo {pages} {607} (\bibinfo {year}
  {2020})}\BibitemShut {NoStop}%
\bibitem [{\citenamefont {Han}\ \emph {et~al.}(2014)\citenamefont {Han},
  \citenamefont {Westley},\ and\ \citenamefont {Sen}}]{Han2014}%
  \BibitemOpen
  \bibfield  {author} {\bibinfo {author} {\bibfnamefont {D.}~\bibnamefont
  {Han}}, \bibinfo {author} {\bibfnamefont {M.}~\bibnamefont {Westley}}, \ and\
  \bibinfo {author} {\bibfnamefont {S.}~\bibnamefont {Sen}},\ }\href {\doibase
  10.1103/PhysRevE.90.032904} {\bibfield  {journal} {\bibinfo  {journal}
  {Physical Review E - Statistical, Nonlinear, and Soft Matter Physics}\
  }\textbf {\bibinfo {volume} {90}},\ \bibinfo {pages} {1} (\bibinfo {year}
  {2014})}\BibitemShut {NoStop}%
\bibitem [{\citenamefont {Kashyap}\ and\ \citenamefont {Sen}(2021)}]{sen2}%
  \BibitemOpen
  \bibfield  {author} {\bibinfo {author} {\bibfnamefont {R.}~\bibnamefont
  {Kashyap}}\ and\ \bibinfo {author} {\bibfnamefont {S.}~\bibnamefont {Sen}},\
  }\href {https://arxiv.org/abs/2105.05028} {\bibfield  {journal} {\bibinfo
  {journal} {arXiv preprint arXiv:2105.05028}\ } (\bibinfo {year}
  {2021})}\BibitemShut {NoStop}%
\bibitem [{\citenamefont {Merkel}\ \emph {et~al.}(2011)\citenamefont {Merkel},
  \citenamefont {Tournat},\ and\ \citenamefont {Gusev}}]{Merkel}%
  \BibitemOpen
  \bibfield  {author} {\bibinfo {author} {\bibfnamefont {A.}~\bibnamefont
  {Merkel}}, \bibinfo {author} {\bibfnamefont {V.}~\bibnamefont {Tournat}}, \
  and\ \bibinfo {author} {\bibfnamefont {V.}~\bibnamefont {Gusev}},\ }\href
  {\doibase 10.1103/PhysRevLett.107.225502} {\bibfield  {journal} {\bibinfo
  {journal} {Physical Review Letters}\ }\textbf {\bibinfo {volume} {107}}
  (\bibinfo {year} {2011}),\ 10.1103/PhysRevLett.107.225502}\BibitemShut
  {NoStop}%
\bibitem [{\citenamefont {Pichard}\ \emph {et~al.}(2014)\citenamefont
  {Pichard}, \citenamefont {Duclos}, \citenamefont {Groby}, \citenamefont
  {Tournat},\ and\ \citenamefont {Gusev}}]{Pichard2014}%
  \BibitemOpen
  \bibfield  {author} {\bibinfo {author} {\bibfnamefont {H.}~\bibnamefont
  {Pichard}}, \bibinfo {author} {\bibfnamefont {A.}~\bibnamefont {Duclos}},
  \bibinfo {author} {\bibfnamefont {J.~P.}\ \bibnamefont {Groby}}, \bibinfo
  {author} {\bibfnamefont {V.}~\bibnamefont {Tournat}}, \ and\ \bibinfo
  {author} {\bibfnamefont {V.~E.}\ \bibnamefont {Gusev}},\ }\href {\doibase
  10.1103/PhysRevE.89.013201} {\bibfield  {journal} {\bibinfo  {journal}
  {Physical Review E - Statistical, Nonlinear, and Soft Matter Physics}\
  }\textbf {\bibinfo {volume} {89}},\ \bibinfo {pages} {13201} (\bibinfo {year}
  {2014})}\BibitemShut {NoStop}%
\bibitem [{\citenamefont {K{\"{o}}pfler}\ \emph {et~al.}(2019)\citenamefont
  {K{\"{o}}pfler}, \citenamefont {Frenzel}, \citenamefont {Kadic},
  \citenamefont {Schmalian},\ and\ \citenamefont {Wegener}}]{Kopfler2019}%
  \BibitemOpen
  \bibfield  {author} {\bibinfo {author} {\bibfnamefont {J.}~\bibnamefont
  {K{\"{o}}pfler}}, \bibinfo {author} {\bibfnamefont {T.}~\bibnamefont
  {Frenzel}}, \bibinfo {author} {\bibfnamefont {M.}~\bibnamefont {Kadic}},
  \bibinfo {author} {\bibfnamefont {J.}~\bibnamefont {Schmalian}}, \ and\
  \bibinfo {author} {\bibfnamefont {M.}~\bibnamefont {Wegener}},\ }\href
  {\doibase 10.1103/PhysRevApplied.11.034059} {\bibfield  {journal} {\bibinfo
  {journal} {Physical Review Applied}\ }\textbf {\bibinfo {volume} {11}},\
  \bibinfo {pages} {34059} (\bibinfo {year} {2019})}\BibitemShut {NoStop}%
\bibitem [{\citenamefont {Ngapasare}\ \emph {et~al.}(2020)\citenamefont
  {Ngapasare}, \citenamefont {Theocharis}, \citenamefont {Richoux},
  \citenamefont {Skokos},\ and\ \citenamefont {Achilleos}}]{Ngapasare2020}%
  \BibitemOpen
  \bibfield  {author} {\bibinfo {author} {\bibfnamefont {A.}~\bibnamefont
  {Ngapasare}}, \bibinfo {author} {\bibfnamefont {G.}~\bibnamefont
  {Theocharis}}, \bibinfo {author} {\bibfnamefont {O.}~\bibnamefont {Richoux}},
  \bibinfo {author} {\bibfnamefont {C.}~\bibnamefont {Skokos}}, \ and\ \bibinfo
  {author} {\bibfnamefont {V.}~\bibnamefont {Achilleos}},\ }\href {\doibase
  10.1103/PhysRevB.102.054201} {\bibfield  {journal} {\bibinfo  {journal}
  {Physical Review B}\ }\textbf {\bibinfo {volume} {102}},\ \bibinfo {pages}
  {54201} (\bibinfo {year} {2020})}\BibitemShut {NoStop}%
\bibitem [{\citenamefont {Allein}\ \emph {et~al.}(2017)\citenamefont {Allein},
  \citenamefont {Tournat}, \citenamefont {Gusev},\ and\ \citenamefont
  {Theocharis}}]{Allein2017}%
  \BibitemOpen
  \bibfield  {author} {\bibinfo {author} {\bibfnamefont {F.}~\bibnamefont
  {Allein}}, \bibinfo {author} {\bibfnamefont {V.}~\bibnamefont {Tournat}},
  \bibinfo {author} {\bibfnamefont {V.~E.}\ \bibnamefont {Gusev}}, \ and\
  \bibinfo {author} {\bibfnamefont {G.}~\bibnamefont {Theocharis}},\ }\href
  {\doibase 10.1016/j.eml.2016.08.001} {\bibfield  {journal} {\bibinfo
  {journal} {Extreme Mechanics Letters}\ }\textbf {\bibinfo {volume} {12}},\
  \bibinfo {pages} {65} (\bibinfo {year} {2017})},\ \Eprint
  {http://arxiv.org/abs/1607.00831} {arXiv:1607.00831} \BibitemShut {NoStop}%
\bibitem [{\citenamefont {Dubus}\ \emph {et~al.}(2016)\citenamefont {Dubus},
  \citenamefont {Swinteck}, \citenamefont {Muralidharan}, \citenamefont
  {Vasseur},\ and\ \citenamefont {Deymier}}]{Dubus2016}%
  \BibitemOpen
  \bibfield  {author} {\bibinfo {author} {\bibfnamefont {B.}~\bibnamefont
  {Dubus}}, \bibinfo {author} {\bibfnamefont {N.}~\bibnamefont {Swinteck}},
  \bibinfo {author} {\bibfnamefont {K.}~\bibnamefont {Muralidharan}}, \bibinfo
  {author} {\bibfnamefont {J.~O.}\ \bibnamefont {Vasseur}}, \ and\ \bibinfo
  {author} {\bibfnamefont {P.~A.}\ \bibnamefont {Deymier}},\ }\href {\doibase
  10.1115/1.4033457} {\bibfield  {journal} {\bibinfo  {journal} {Journal of
  Vibration and Acoustics, Transactions of the ASME}\ }\textbf {\bibinfo
  {volume} {138}} (\bibinfo {year} {2016}),\ 10.1115/1.4033457}\BibitemShut
  {NoStop}%
\bibitem [{\citenamefont {Deng}\ \emph {et~al.}(2018)\citenamefont {Deng},
  \citenamefont {Wang}, \citenamefont {He}, \citenamefont {Tournat},\ and\
  \citenamefont {Bertoldi}}]{Deng2018}%
  \BibitemOpen
  \bibfield  {author} {\bibinfo {author} {\bibfnamefont {B.}~\bibnamefont
  {Deng}}, \bibinfo {author} {\bibfnamefont {P.}~\bibnamefont {Wang}}, \bibinfo
  {author} {\bibfnamefont {Q.}~\bibnamefont {He}}, \bibinfo {author}
  {\bibfnamefont {V.}~\bibnamefont {Tournat}}, \ and\ \bibinfo {author}
  {\bibfnamefont {K.}~\bibnamefont {Bertoldi}},\ }\href {\doibase
  10.1038/s41467-018-05908-9} {\bibfield  {journal} {\bibinfo  {journal}
  {Nature Communications}\ }\textbf {\bibinfo {volume} {9}},\ \bibinfo {pages}
  {1} (\bibinfo {year} {2018})}\BibitemShut {NoStop}%
\bibitem [{\citenamefont {Zhang}\ \emph {et~al.}(2019)\citenamefont {Zhang},
  \citenamefont {Umnova},\ and\ \citenamefont {Venegas}}]{Zhang2019}%
  \BibitemOpen
  \bibfield  {author} {\bibinfo {author} {\bibfnamefont {Q.}~\bibnamefont
  {Zhang}}, \bibinfo {author} {\bibfnamefont {O.}~\bibnamefont {Umnova}}, \
  and\ \bibinfo {author} {\bibfnamefont {R.}~\bibnamefont {Venegas}},\ }\href
  {\doibase 10.1103/PhysRevE.100.062206} {\bibfield  {journal} {\bibinfo
  {journal} {Physical Review E}\ }\textbf {\bibinfo {volume} {100}},\ \bibinfo
  {pages} {062206} (\bibinfo {year} {2019})}\BibitemShut {NoStop}%
\bibitem [{\citenamefont {Yasuda}\ \emph {et~al.}(2019)\citenamefont {Yasuda},
  \citenamefont {Miyazawa}, \citenamefont {Charalampidis}, \citenamefont
  {Chong}, \citenamefont {Kevrekidis},\ and\ \citenamefont
  {Yang}}]{Yasuda2019}%
  \BibitemOpen
  \bibfield  {author} {\bibinfo {author} {\bibfnamefont {H.}~\bibnamefont
  {Yasuda}}, \bibinfo {author} {\bibfnamefont {Y.}~\bibnamefont {Miyazawa}},
  \bibinfo {author} {\bibfnamefont {E.~G.}\ \bibnamefont {Charalampidis}},
  \bibinfo {author} {\bibfnamefont {C.}~\bibnamefont {Chong}}, \bibinfo
  {author} {\bibfnamefont {P.~G.}\ \bibnamefont {Kevrekidis}}, \ and\ \bibinfo
  {author} {\bibfnamefont {J.}~\bibnamefont {Yang}},\ }\href {\doibase
  10.1126/sciadv.aau2835} {\bibfield  {journal} {\bibinfo  {journal} {Science
  Advances}\ }\textbf {\bibinfo {volume} {5}},\ \bibinfo {pages} {1} (\bibinfo
  {year} {2019})},\ \Eprint {http://arxiv.org/abs/1805.05909}
  {arXiv:1805.05909} \BibitemShut {NoStop}%
\bibitem [{\citenamefont {Deng}\ \emph {et~al.}(2019)\citenamefont {Deng},
  \citenamefont {Mo}, \citenamefont {Tournat}, \citenamefont {Bertoldi},\ and\
  \citenamefont {Raney}}]{Deng2019}%
  \BibitemOpen
  \bibfield  {author} {\bibinfo {author} {\bibfnamefont {B.}~\bibnamefont
  {Deng}}, \bibinfo {author} {\bibfnamefont {C.}~\bibnamefont {Mo}}, \bibinfo
  {author} {\bibfnamefont {V.}~\bibnamefont {Tournat}}, \bibinfo {author}
  {\bibfnamefont {K.}~\bibnamefont {Bertoldi}}, \ and\ \bibinfo {author}
  {\bibfnamefont {J.~R.}\ \bibnamefont {Raney}},\ }\href {\doibase
  10.1103/PhysRevLett.123.024101} {\bibfield  {journal} {\bibinfo  {journal}
  {Physical Review Letters}\ }\textbf {\bibinfo {volume} {123}},\ \bibinfo
  {pages} {24101} (\bibinfo {year} {2019})}\BibitemShut {NoStop}%
\bibitem [{\citenamefont {Sugino}\ \emph {et~al.}(2017)\citenamefont {Sugino},
  \citenamefont {Xia}, \citenamefont {Leadenham}, \citenamefont {Ruzzene},\
  and\ \citenamefont {Erturk}}]{Sugino2017}%
  \BibitemOpen
  \bibfield  {author} {\bibinfo {author} {\bibfnamefont {C.}~\bibnamefont
  {Sugino}}, \bibinfo {author} {\bibfnamefont {Y.}~\bibnamefont {Xia}},
  \bibinfo {author} {\bibfnamefont {S.}~\bibnamefont {Leadenham}}, \bibinfo
  {author} {\bibfnamefont {M.}~\bibnamefont {Ruzzene}}, \ and\ \bibinfo
  {author} {\bibfnamefont {A.}~\bibnamefont {Erturk}},\ }\href {\doibase
  10.1016/j.jsv.2017.06.004} {\bibfield  {journal} {\bibinfo  {journal}
  {Journal of Sound and Vibration}\ }\textbf {\bibinfo {volume} {406}},\
  \bibinfo {pages} {104} (\bibinfo {year} {2017})},\ \Eprint
  {http://arxiv.org/abs/1612.03130} {arXiv:1612.03130} \BibitemShut {NoStop}%
\bibitem [{\citenamefont {Beli}\ \emph {et~al.}(2018)\citenamefont {Beli},
  \citenamefont {Arruda},\ and\ \citenamefont {Ruzzene}}]{Beli2018}%
  \BibitemOpen
  \bibfield  {author} {\bibinfo {author} {\bibfnamefont {D.}~\bibnamefont
  {Beli}}, \bibinfo {author} {\bibfnamefont {J.~R.}\ \bibnamefont {Arruda}}, \
  and\ \bibinfo {author} {\bibfnamefont {M.}~\bibnamefont {Ruzzene}},\ }\href
  {\doibase 10.1016/j.ijsolstr.2018.01.027} {\bibfield  {journal} {\bibinfo
  {journal} {International Journal of Solids and Structures}\ }\textbf
  {\bibinfo {volume} {139-140}},\ \bibinfo {pages} {105} (\bibinfo {year}
  {2018})}\BibitemShut {NoStop}%
\bibitem [{\citenamefont {Karttunen}\ and\ \citenamefont
  {Reddy}(2020)}]{Karttunen2020}%
  \BibitemOpen
  \bibfield  {author} {\bibinfo {author} {\bibfnamefont {A.~T.}\ \bibnamefont
  {Karttunen}}\ and\ \bibinfo {author} {\bibfnamefont {J.~N.}\ \bibnamefont
  {Reddy}},\ }\href {\doibase 10.1016/j.ijsolstr.2020.08.020} {\bibfield
  {journal} {\bibinfo  {journal} {International Journal of Solids and
  Structures}\ }\textbf {\bibinfo {volume} {204-205}},\ \bibinfo {pages} {172}
  (\bibinfo {year} {2020})}\BibitemShut {NoStop}%
\bibitem [{\citenamefont {Wang}\ \emph {et~al.}(2018)\citenamefont {Wang},
  \citenamefont {Liu}, \citenamefont {Zhu},\ and\ \citenamefont
  {Hu}}]{Wang2018a}%
  \BibitemOpen
  \bibfield  {author} {\bibinfo {author} {\bibfnamefont {Y.~T.}\ \bibnamefont
  {Wang}}, \bibinfo {author} {\bibfnamefont {X.~N.}\ \bibnamefont {Liu}},
  \bibinfo {author} {\bibfnamefont {R.}~\bibnamefont {Zhu}}, \ and\ \bibinfo
  {author} {\bibfnamefont {G.~K.}\ \bibnamefont {Hu}},\ }\href {\doibase
  10.1038/s41598-018-29816-6} {\bibfield  {journal} {\bibinfo  {journal}
  {Scientific Reports}\ }\textbf {\bibinfo {volume} {8}},\ \bibinfo {pages} {1}
  (\bibinfo {year} {2018})}\BibitemShut {NoStop}%
\bibitem [{\citenamefont {Yin}\ \emph {et~al.}(2020)\citenamefont {Yin},
  \citenamefont {Zhang}, \citenamefont {Xu}, \citenamefont {Zhang},\ and\
  \citenamefont {Gao}}]{Yin2020}%
  \BibitemOpen
  \bibfield  {author} {\bibinfo {author} {\bibfnamefont {X.}~\bibnamefont
  {Yin}}, \bibinfo {author} {\bibfnamefont {S.}~\bibnamefont {Zhang}}, \bibinfo
  {author} {\bibfnamefont {G.-K.}\ \bibnamefont {Xu}}, \bibinfo {author}
  {\bibfnamefont {L.-Y.}\ \bibnamefont {Zhang}}, \ and\ \bibinfo {author}
  {\bibfnamefont {Z.-Y.}\ \bibnamefont {Gao}},\ }\href {\doibase
  10.1016/j.eml.2020.100668} {\bibfield  {journal} {\bibinfo  {journal}
  {Extreme Mechanics Letters}\ }\textbf {\bibinfo {volume} {36}},\ \bibinfo
  {pages} {100668} (\bibinfo {year} {2020})}\BibitemShut {NoStop}%
\bibitem [{\citenamefont {Zhang}\ \emph {et~al.}(2021)\citenamefont {Zhang},
  \citenamefont {Yin}, \citenamefont {Yang}, \citenamefont {Li},\ and\
  \citenamefont {Xu}}]{Zhang2021}%
  \BibitemOpen
  \bibfield  {author} {\bibinfo {author} {\bibfnamefont {L.~Y.}\ \bibnamefont
  {Zhang}}, \bibinfo {author} {\bibfnamefont {X.}~\bibnamefont {Yin}}, \bibinfo
  {author} {\bibfnamefont {J.}~\bibnamefont {Yang}}, \bibinfo {author}
  {\bibfnamefont {A.}~\bibnamefont {Li}}, \ and\ \bibinfo {author}
  {\bibfnamefont {G.~K.}\ \bibnamefont {Xu}},\ }\href {\doibase
  10.1016/j.compscitech.2021.108740} {\bibfield  {journal} {\bibinfo  {journal}
  {Composites Science and Technology}\ }\textbf {\bibinfo {volume} {207}},\
  \bibinfo {pages} {108740} (\bibinfo {year} {2021})}\BibitemShut {NoStop}%
\bibitem [{\citenamefont {Fang}\ \emph {et~al.}(2020)\citenamefont {Fang},
  \citenamefont {Chang},\ and\ \citenamefont {Wang}}]{Fang2020}%
  \BibitemOpen
  \bibfield  {author} {\bibinfo {author} {\bibfnamefont {H.}~\bibnamefont
  {Fang}}, \bibinfo {author} {\bibfnamefont {T.~S.}\ \bibnamefont {Chang}}, \
  and\ \bibinfo {author} {\bibfnamefont {K.~W.}\ \bibnamefont {Wang}},\ }\href
  {\doibase 10.1088/1361-665X/ab524e} {\bibfield  {journal} {\bibinfo
  {journal} {Smart Materials and Structures}\ }\textbf {\bibinfo {volume} {29}}
  (\bibinfo {year} {2020}),\ 10.1088/1361-665X/ab524e}\BibitemShut {NoStop}%
\bibitem [{\citenamefont {Pratapa}\ \emph {et~al.}(2018)\citenamefont
  {Pratapa}, \citenamefont {Suryanarayana},\ and\ \citenamefont
  {Paulino}}]{Pratapa2018}%
  \BibitemOpen
  \bibfield  {author} {\bibinfo {author} {\bibfnamefont {P.~P.}\ \bibnamefont
  {Pratapa}}, \bibinfo {author} {\bibfnamefont {P.}~\bibnamefont
  {Suryanarayana}}, \ and\ \bibinfo {author} {\bibfnamefont {G.~H.}\
  \bibnamefont {Paulino}},\ }\href {\doibase 10.1016/j.jmps.2018.05.012}
  {\bibfield  {journal} {\bibinfo  {journal} {Journal of the Mechanics and
  Physics of Solids}\ }\textbf {\bibinfo {volume} {118}},\ \bibinfo {pages}
  {115} (\bibinfo {year} {2018})}\BibitemShut {NoStop}%
\bibitem [{\citenamefont {Peyrard}\ and\ \citenamefont
  {Bishop}(1989)}]{Peyrard1989}%
  \BibitemOpen
  \bibfield  {author} {\bibinfo {author} {\bibfnamefont {M.}~\bibnamefont
  {Peyrard}}\ and\ \bibinfo {author} {\bibfnamefont {A.~R.}\ \bibnamefont
  {Bishop}},\ }\href {\doibase 10.1103/PhysRevLett.62.2755} {\bibfield
  {journal} {\bibinfo  {journal} {Physical Review Letters}\ }\textbf {\bibinfo
  {volume} {62}},\ \bibinfo {pages} {2755} (\bibinfo {year}
  {1989})}\BibitemShut {NoStop}%
\bibitem [{\citenamefont {Dauxois}\ \emph {et~al.}(1993)\citenamefont
  {Dauxois}, \citenamefont {Peyrard},\ and\ \citenamefont
  {Bishop}}]{Dauxois1993}%
  \BibitemOpen
  \bibfield  {author} {\bibinfo {author} {\bibfnamefont {T.}~\bibnamefont
  {Dauxois}}, \bibinfo {author} {\bibfnamefont {M.}~\bibnamefont {Peyrard}}, \
  and\ \bibinfo {author} {\bibfnamefont {A.~R.}\ \bibnamefont {Bishop}},\
  }\href {\doibase 10.1103/PhysRevE.47.684} {\bibfield  {journal} {\bibinfo
  {journal} {Physical Review E}\ }\textbf {\bibinfo {volume} {47}},\ \bibinfo
  {pages} {684} (\bibinfo {year} {1993})}\BibitemShut {NoStop}%
\bibitem [{\citenamefont {Chevizovich}\ \emph {et~al.}(2020)\citenamefont
  {Chevizovich}, \citenamefont {Michieletto}, \citenamefont {Mvogo},
  \citenamefont {Zakiryanov},\ and\ \citenamefont
  {Zdravkovi{\'{c}}}}]{Chevizovich2020}%
  \BibitemOpen
  \bibfield  {author} {\bibinfo {author} {\bibfnamefont {D.}~\bibnamefont
  {Chevizovich}}, \bibinfo {author} {\bibfnamefont {D.}~\bibnamefont
  {Michieletto}}, \bibinfo {author} {\bibfnamefont {A.}~\bibnamefont {Mvogo}},
  \bibinfo {author} {\bibfnamefont {F.}~\bibnamefont {Zakiryanov}}, \ and\
  \bibinfo {author} {\bibfnamefont {S.}~\bibnamefont {Zdravkovi{\'{c}}}},\
  }\href {\doibase 10.1098/rsos.200774} {\bibfield  {journal} {\bibinfo
  {journal} {Royal Society Open Science}\ }\textbf {\bibinfo {volume} {7}},\
  \bibinfo {pages} {200774} (\bibinfo {year} {2020})}\BibitemShut {NoStop}%
\bibitem [{\citenamefont {Kresling}(2012)}]{Kresling2012}%
  \BibitemOpen
  \bibfield  {author} {\bibinfo {author} {\bibfnamefont {B.}~\bibnamefont
  {Kresling}},\ }in\ \href {\doibase 10.1557/opl.2012.536} {\emph {\bibinfo
  {booktitle} {Materials Research Society Symposium Proceedings}}},\ Vol.\
  \bibinfo {volume} {1420}\ (\bibinfo {year} {2012})\ pp.\ \bibinfo {pages}
  {42--54}\BibitemShut {NoStop}%
\bibitem [{\citenamefont {Howell}\ \emph {et~al.}(2013)\citenamefont {Howell},
  \citenamefont {Magleby},\ and\ \citenamefont {Olsen}}]{Howell2013}%
  \BibitemOpen
  \bibfield  {author} {\bibinfo {author} {\bibfnamefont {L.~L.}\ \bibnamefont
  {Howell}}, \bibinfo {author} {\bibfnamefont {S.~P.}\ \bibnamefont {Magleby}},
  \ and\ \bibinfo {author} {\bibfnamefont {B.~M.}\ \bibnamefont {Olsen}},\
  }\href {\doibase 10.1002/9781118516485} {\emph {\bibinfo {title} {Handbook of
  Compliant Mechanisms}}}\ (\bibinfo  {publisher} {John Wiley and Sons},\
  \bibinfo {year} {2013})\BibitemShut {NoStop}%
\bibitem [{\citenamefont {Nesterenko}(2001)}]{Nesterenko2001}%
  \BibitemOpen
  \bibfield  {author} {\bibinfo {author} {\bibfnamefont {V.~F.}\ \bibnamefont
  {Nesterenko}},\ }\href {\doibase 10.1007/978-1-4757-3524-6} {\emph {\bibinfo
  {title} {Dynamics of Heterogeneous Materials}}}\ (\bibinfo  {publisher}
  {Springer New York},\ \bibinfo {year} {2001})\BibitemShut {NoStop}%
\bibitem [{\citenamefont {Huang}\ \emph {et~al.}(1993)\citenamefont {Huang},
  \citenamefont {Shi},\ and\ \citenamefont {Xu}}]{Huang1993}%
  \BibitemOpen
  \bibfield  {author} {\bibinfo {author} {\bibfnamefont {G.}~\bibnamefont
  {Huang}}, \bibinfo {author} {\bibfnamefont {Z.~P.}\ \bibnamefont {Shi}}, \
  and\ \bibinfo {author} {\bibfnamefont {Z.}~\bibnamefont {Xu}},\ }\href
  {\doibase 10.1103/PhysRevB.47.14561} {\bibfield  {journal} {\bibinfo
  {journal} {Physical Review B}\ }\textbf {\bibinfo {volume} {47}},\ \bibinfo
  {pages} {14561} (\bibinfo {year} {1993})}\BibitemShut {NoStop}%
\bibitem [{\citenamefont {Chong}\ and\ \citenamefont
  {Kevrekidis}(2018)}]{Chong2018}%
  \BibitemOpen
  \bibfield  {author} {\bibinfo {author} {\bibfnamefont {C.}~\bibnamefont
  {Chong}}\ and\ \bibinfo {author} {\bibfnamefont {P.~G.}\ \bibnamefont
  {Kevrekidis}},\ }\href@noop {} {\emph {\bibinfo {title} {{Coherent Structures
  in Granular Crystals: From Experiment and Modelling to Computation and
  Mathematical Analysis}}}},\ \bibinfo {edition} {1st}\ ed.\ (\bibinfo
  {publisher} {Springer International Publishing},\ \bibinfo {year}
  {2018})\BibitemShut {NoStop}%
\bibitem [{\citenamefont {Manakov}(1973)}]{Manakov1974}%
  \BibitemOpen
  \bibfield  {author} {\bibinfo {author} {\bibfnamefont {S.}~\bibnamefont
  {Manakov}},\ }\href@noop {} {\bibfield  {journal} {\bibinfo  {journal}
  {Journal of Experimental and Theoretical Physics}\ }\textbf {\bibinfo
  {volume} {38}},\ \bibinfo {pages} {248} (\bibinfo {year} {1973})}\BibitemShut
  {NoStop}%
\bibitem [{\citenamefont {Haelterman}\ \emph {et~al.}(1993)\citenamefont
  {Haelterman}, \citenamefont {Sheppard},\ and\ \citenamefont
  {Snyder}}]{Haelterman1993}%
  \BibitemOpen
  \bibfield  {author} {\bibinfo {author} {\bibfnamefont {M.}~\bibnamefont
  {Haelterman}}, \bibinfo {author} {\bibfnamefont {A.~P.}\ \bibnamefont
  {Sheppard}}, \ and\ \bibinfo {author} {\bibfnamefont {A.~W.}\ \bibnamefont
  {Snyder}},\ }\href {\doibase 10.1364/ol.18.001406} {\bibfield  {journal}
  {\bibinfo  {journal} {Optics Letters}\ }\textbf {\bibinfo {volume} {18}},\
  \bibinfo {pages} {1406} (\bibinfo {year} {1993})}\BibitemShut {NoStop}%
\bibitem [{\citenamefont {Haelterman}\ and\ \citenamefont
  {Sheppard}(1994{\natexlab{a}})}]{Haelterman1994}%
  \BibitemOpen
  \bibfield  {author} {\bibinfo {author} {\bibfnamefont {M.}~\bibnamefont
  {Haelterman}}\ and\ \bibinfo {author} {\bibfnamefont {A.~P.}\ \bibnamefont
  {Sheppard}},\ }\href {\doibase 10.1016/0375-9601(94)91282-3} {\bibfield
  {journal} {\bibinfo  {journal} {Physics Letters A}\ }\textbf {\bibinfo
  {volume} {194}},\ \bibinfo {pages} {191} (\bibinfo {year}
  {1994}{\natexlab{a}})}\BibitemShut {NoStop}%
\bibitem [{\citenamefont {Haelterman}\ and\ \citenamefont
  {Sheppard}(1994{\natexlab{b}})}]{Haelterman1994a}%
  \BibitemOpen
  \bibfield  {author} {\bibinfo {author} {\bibfnamefont {M.}~\bibnamefont
  {Haelterman}}\ and\ \bibinfo {author} {\bibfnamefont {A.}~\bibnamefont
  {Sheppard}},\ }\href {\doibase 10.1103/PhysRevE.49.3376} {\bibfield
  {journal} {\bibinfo  {journal} {Physical Review E}\ }\textbf {\bibinfo
  {volume} {49}},\ \bibinfo {pages} {3376} (\bibinfo {year}
  {1994}{\natexlab{b}})}\BibitemShut {NoStop}%
\bibitem [{\citenamefont {Stalin}\ \emph {et~al.}(2019)\citenamefont {Stalin},
  \citenamefont {Ramakrishnan}, \citenamefont {Senthilvelan},\ and\
  \citenamefont {Lakshmanan}}]{Stalin2019}%
  \BibitemOpen
  \bibfield  {author} {\bibinfo {author} {\bibfnamefont {S.}~\bibnamefont
  {Stalin}}, \bibinfo {author} {\bibfnamefont {R.}~\bibnamefont
  {Ramakrishnan}}, \bibinfo {author} {\bibfnamefont {M.}~\bibnamefont
  {Senthilvelan}}, \ and\ \bibinfo {author} {\bibfnamefont {M.}~\bibnamefont
  {Lakshmanan}},\ }\href {\doibase 10.1103/PhysRevLett.122.043901} {\bibfield
  {journal} {\bibinfo  {journal} {Physical Review Letters}\ }\textbf {\bibinfo
  {volume} {122}} (\bibinfo {year} {2019}),\ 10.1103/PhysRevLett.122.043901},\
  \Eprint {http://arxiv.org/abs/1810.01331} {arXiv:1810.01331} \BibitemShut
  {NoStop}%
\bibitem [{\citenamefont {Ramakrishnan}\ \emph {et~al.}(2020)\citenamefont
  {Ramakrishnan}, \citenamefont {Stalin},\ and\ \citenamefont
  {Lakshmanan}}]{Ramakrishnan2020}%
  \BibitemOpen
  \bibfield  {author} {\bibinfo {author} {\bibfnamefont {R.}~\bibnamefont
  {Ramakrishnan}}, \bibinfo {author} {\bibfnamefont {S.}~\bibnamefont
  {Stalin}}, \ and\ \bibinfo {author} {\bibfnamefont {M.}~\bibnamefont
  {Lakshmanan}},\ }\href {\doibase 10.1103/PhysRevE.102.042212} {\bibfield
  {journal} {\bibinfo  {journal} {Physical Review E}\ }\textbf {\bibinfo
  {volume} {102}} (\bibinfo {year} {2020}),\
  10.1103/PhysRevE.102.042212}\BibitemShut {NoStop}%
\bibitem [{\citenamefont {Kivshar}\ and\ \citenamefont
  {Agrawal}(2003)}]{Kivshar2003}%
  \BibitemOpen
  \bibfield  {author} {\bibinfo {author} {\bibfnamefont {Y.~S.}\ \bibnamefont
  {Kivshar}}\ and\ \bibinfo {author} {\bibfnamefont {G.~P.}\ \bibnamefont
  {Agrawal}},\ }\href {\doibase 10.1016/B978-0-12-410590-4.X5000-1} {\emph
  {\bibinfo {title} {Optical Solitons: From Fibers to Photonic Crystals}}}\
  (\bibinfo  {publisher} {Academic Press},\ \bibinfo {year} {2003})\ pp.\
  \bibinfo {pages} {1--540}\BibitemShut {NoStop}%
\bibitem [{\citenamefont {Baronio}\ \emph {et~al.}(2012)\citenamefont
  {Baronio}, \citenamefont {Degasperis}, \citenamefont {Conforti},\ and\
  \citenamefont {Wabnitz}}]{Baronio2012}%
  \BibitemOpen
  \bibfield  {author} {\bibinfo {author} {\bibfnamefont {F.}~\bibnamefont
  {Baronio}}, \bibinfo {author} {\bibfnamefont {A.}~\bibnamefont {Degasperis}},
  \bibinfo {author} {\bibfnamefont {M.}~\bibnamefont {Conforti}}, \ and\
  \bibinfo {author} {\bibfnamefont {S.}~\bibnamefont {Wabnitz}},\ }\href
  {\doibase 10.1103/PhysRevLett.109.044102} {\bibfield  {journal} {\bibinfo
  {journal} {Physical Review Letters}\ }\textbf {\bibinfo {volume} {109}}
  (\bibinfo {year} {2012}),\ 10.1103/PhysRevLett.109.044102}\BibitemShut
  {NoStop}%
\bibitem [{\citenamefont {Bertola}\ and\ \citenamefont
  {Tovbis}(2013)}]{Bertola2013}%
  \BibitemOpen
  \bibfield  {author} {\bibinfo {author} {\bibfnamefont {M.}~\bibnamefont
  {Bertola}}\ and\ \bibinfo {author} {\bibfnamefont {A.}~\bibnamefont
  {Tovbis}},\ }\href {\doibase 10.1002/cpa.21445} {\bibfield  {journal}
  {\bibinfo  {journal} {Communications on Pure and Applied Mathematics}\
  }\textbf {\bibinfo {volume} {66}},\ \bibinfo {pages} {678} (\bibinfo {year}
  {2013})},\ \Eprint {http://arxiv.org/abs/1004.1828} {arXiv:1004.1828}
  \BibitemShut {NoStop}%
\bibitem [{\citenamefont {Chong}\ \emph {et~al.}(2021)\citenamefont {Chong},
  \citenamefont {Wang}, \citenamefont {Mar{\'{e}}chal}, \citenamefont
  {Charalampidis}, \citenamefont {Moler{\'{o}}n}, \citenamefont
  {Mart{\'{i}}nez}, \citenamefont {Porter}, \citenamefont {Kevrekidis},\ and\
  \citenamefont {Daraio}}]{Chong_2021}%
  \BibitemOpen
  \bibfield  {author} {\bibinfo {author} {\bibfnamefont {C.}~\bibnamefont
  {Chong}}, \bibinfo {author} {\bibfnamefont {Y.}~\bibnamefont {Wang}},
  \bibinfo {author} {\bibfnamefont {D.}~\bibnamefont {Mar{\'{e}}chal}},
  \bibinfo {author} {\bibfnamefont {E.~G.}\ \bibnamefont {Charalampidis}},
  \bibinfo {author} {\bibfnamefont {M.}~\bibnamefont {Moler{\'{o}}n}}, \bibinfo
  {author} {\bibfnamefont {A.~J.}\ \bibnamefont {Mart{\'{i}}nez}}, \bibinfo
  {author} {\bibfnamefont {M.~A.}\ \bibnamefont {Porter}}, \bibinfo {author}
  {\bibfnamefont {P.~G.}\ \bibnamefont {Kevrekidis}}, \ and\ \bibinfo {author}
  {\bibfnamefont {C.}~\bibnamefont {Daraio}},\ }\href {\doibase
  10.1088/1367-2630/abdb6f} {\bibfield  {journal} {\bibinfo  {journal} {New
  Journal of Physics}\ }\textbf {\bibinfo {volume} {23}},\ \bibinfo {pages}
  {43008} (\bibinfo {year} {2021})},\ \Eprint {http://arxiv.org/abs/2009.10300}
  {arXiv:2009.10300} \BibitemShut {NoStop}%
\end{thebibliography}%
\end{document}